\documentclass[a4paper,11pt]{article}

\pdfoutput=1 

\usepackage{jcappub} 

\usepackage{amsmath, amssymb, amsthm, graphicx, epsfig, fancyhdr,epsfig, slashed, mathrsfs}

\usepackage{tikzsymbols}
\usepackage{float}
\usepackage[T1]{fontenc} 
\usepackage{color}

\definecolor{revisionorange}{HTML}{D95F02}

\def\bl#1\el{\begin{align}#1\end{align}}
\def\ba#1\ea{\begin{align*}#1\end{align*}}

\usepackage{tikz,xcolor,hyperref}

\definecolor{lime}{HTML}{A6CE39}
\DeclareRobustCommand{\orcidicon}{
	\begin{tikzpicture}
	\draw[lime, fill=lime] (0,0) 
	circle [radius=0.2] 
	node[white] {{\fontfamily{qag}\selectfont \tiny ID}};
	\draw[white, fill=white] (-0.0625,0.095) 
	circle [radius=0.007];
	\end{tikzpicture}
	\hspace{-2mm}
}

\foreach \x in {A, ..., Z}{\expandafter\xdef\csname orcid\x\endcsname{\noexpand\href{https://orcid.org/\csname orcidauthor\x\endcsname}
			{\noexpand\orcidicon}}
}

\newcommand{\be}{\begin{equation}}
\newcommand{\ee}{\end{equation}}
\newcommand{\bea}{\begin{eqnarray}}
\newcommand{\eea}{\end{eqnarray}}

\newcommand{\beq}{\begin{equation}}
\newcommand{\eeq}{\end{equation}}

\makeatletter
\gdef\@fpheader{}
\makeatother

\begin{document}

\title{Magnetically assisted primordial scalar perturbations: \\ \it{Scalar-Induced Gravitational Waves}}

\author[a]{Arko Bhaumik,}
\author[b]{Theodoros Papanikolaou,}
\author[c]{Anish Ghoshal}
\affiliation[a]{Physics and Applied Mathematics Unit, Indian Statistical Institute, \\ 203 B.T. Road,
Kolkata 700108, India}

\affiliation[b]{Laboratory of Theoretical and Computational Physics, Department of Physics, University of Patras, 26554, Patras, Greece}

\affiliation[c]{Department of Physics and Astronomy, University of Sussex,
Brighton, BN1 9RH, United Kingdom}

\emailAdd{arkobhaumik12@gmail.com}
\emailAdd{papaniko@upatras.gr}
\emailAdd{a.ghoshal@sussex.ac.uk}

\abstract{Primordial magnetic fields (PMFs) provide a well-motivated source of cosmological perturbations through their anisotropic stress and may leave observable imprints in both the scalar and tensor sectors. In this work, we study scalar metric perturbations sourced by a PMF evolving through a finite post-inflationary epoch characterized by a constant equation-of-state (EoS) parameter $w$. Working in a gauge-invariant framework, we derive the sourced evolution equation for the Bardeen gravitational potential in a general constant-$w$ background. Our analysis indicates faster (slower) growth of PMF-sourced scalar perturbations on superhorizon scales for a stiffer $w>1/3$ (softer $w<1/3$) background compared to the marginal radiation-dominated scenario with $w=1/3$. We then derive the scalar-induced gravitational wave (SIGW) background induced at second order by the magnetically generated scalar perturbations. Our analysis indicates a narrow viable parametric region for a kination-like post-inflationary era and a nearly scale-invariant PMF that may give rise to a PMF-sourced SIGW signal dominating over the direct PMF-generated tensor background. Simultaneously, the induced tensor modes are found to be perturbatively small compared to the magnetic sector, which, in turn, exerts negligible backreaction on the background. Interestingly enough, for an inflationary scale $H_{\rm inf}\sim10^3-10^9$ GeV and duration of reheating in $e$-folds $N_{\rm reh}\sim8-10$, the dominant PMF-sourced SIGW signal may be detectable by next-generation terrestrial and space-based interferometric GW detectors spanning the mHz$-$kHz frequency range.}


\begin{flushright}
\end{flushright}

\maketitle


\section{Introduction}

Cosmic magnetic fields with coherent structure are presently observed across an enormous range of scales from galaxies and galaxy clusters to the intergalactic medium \cite{2004NewAR..48..763V,2011ApJ...728...97V,2019A&A...622A..16O,2010-Neronov.Vovk-Sci,2011ApJ...727L...4D,Tiede:2017aql}, while the existence of such fields strongly motivates the search for a primordial component generated prior to structure formation \cite{TurnerWidrow1988,Ratra1992,Widrow2002,DurrerNeronov2013,Subramanian2016}. Primordial magnetic fields (PMFs) are particularly compelling because they can seed the magnetic fields observed in later cosmic epochs, alter the thermal and ionization history of the plasma \cite{Sethi:2004pe,2012PhRvD..86d3510S} and leave imprints in the cosmic microwave background (CMB) \cite{Durrer:2006pc,2010PhRvD..81d3517S,Kunze:2013uja,Ferreira:2014hma,PlanckPMF2016,Zucca:2016iur,Jedamzik2020}, large-scale structure \cite{Kim:1994zh,Kahniashvili:2012dy,Kunze:2022mlr,Adi:2023doe,Ralegankar:2024arh}, 21 cm and Lyman-$\alpha$ observables \cite{Tashiro:2006uv,Schleicher:2008hc,Kunze:2018cnn,Minoda:2018gxj,Bera:2020jsg,Cruz:2023rmo,Bhaumik:2024efz,Pavicevic:2025gqi,Bhaumik:2026mlc}, being associated as well to stochastic gravitational-wave (GW) backgrounds \cite{Mack2002,Sharma:2019jtb,Brandenburg:2021bfx,Maiti:2024nhv,Atkins:2025pvg,Maiti:2025awl,Bhaumik:2025kuj,Teuscher:2025jhq,Teuscher:2025xke,Maiti:2026hsn,Ragavendra:2026fgs}. At the same time, the very weakness of present observational bounds implies that PMFs, if they exist, must be theoretically modeled in a logically consistent framework in which their generation, subsequent redshifting, and backreaction on the cosmological background are all under quantitative control. A central reason PMFs are cosmologically interesting is that they inevitably carry anisotropic stress. Even when the magnetic energy density constitutes only a subdominant fraction of the total energy budget, the anisotropic stress of the field sources metric perturbations already at linear order and can therefore generate scalar, vector, and tensor fluctuations with distinctive spectral properties \cite{ShawLewis2010,CapriniDurrer2001,Saga2020}. In the scalar sector, the PMF anisotropic stress modifies the evolution of the Bardeen gravitational potentials and can induce a nontrivial evolution of the comoving curvature perturbation on superhorizon scales before neutrino free streaming compensates the magnetic stress \cite{ShawLewis2010}. In the tensor sector, PMFs source a stochastic gravitational-wave background directly at first order, with an amplitude and spectral shape determined by the magnetic power spectrum, the helicity content of the field, and the background expansion history. These effects make PMFs a rare example of an early-Universe sector whose signatures simultaneously involve CMB physics, small-scale primordial perturbations, and a rich GW phenomenology.

An additional layer of interest arises from the rapidly developing theory of GWs induced at second order in cosmological perturbation theory by primordial scalar perturbations. It is by now well understood that scalar fluctuations re-entering the horizon may source a tensor background at quadratic order; this mechanism has become a key probe of  enhanced primordial power on scales far smaller than those accessible to the CMB \cite{Ananda2007,Baumann2007,KohriTerada2018,Domenech:2021ztg}. The scalar-induced gravitational wave (SIGW) background has been studied extensively in connection with inflationary features, spectator fields, phase transitions, and scenarios involving primordial black hole (PBH) formation \cite{SaitoYokoyama2009,Espinosa2018,Cai:2018dig,Bartolo:2018rku,Yuan:2019udt,Pi2019,Domenech:2019quo,Cai:2019jah,Bhattacharya:2019bvk,Domenech:2021ztg,Zhou:2024doz,Maity:2024odg,Kugarajh:2025rbt}. In most of that literature, however, the scalar source is taken to be the primordial curvature perturbation generated during inflation, and/or the post-inflationary background is usually approximated either by instantaneous reheating or by the standard radiation-dominated (RD) evolution. PMFs introduce a qualitatively different possibility: the scalar sector itself can be sourced by the magnetic anisotropic stress, and the resulting magnetically generated scalar perturbations can then act as the seed for a second-order tensor background. This two-step channel is conceptually distinct from the direct first-order PMF tensor signal and deserves to be treated in a dedicated and internally consistent framework.

The post-inflationary expansion history is especially important in this context. Reheating is one of the least constrained stages of early cosmic history, and a broad class of microphysical models can be effectively described by a finite post-inflationary epoch with an approximately constant barotropic equation-of-state (EoS) parameter $w$, bounded by the end of inflation and the onset of RD. The value of $w$ controls the redshifting of the background energy density, the evolution of the comoving Hubble scale, and the time available for superhorizon amplification or subhorizon damping of sourced perturbations \cite{MartinRingeval2010,Cook2015,Dai2014,Domenech2020}. For PMFs, this is particularly consequential, because the magnetic energy density scales as $a^{-4}$ while the background energy density scales as $a^{-3(1+w)}$ during reheating. As a result, the relative importance of the magnetic sector depends sensitively on whether the background fluid is softer or stiffer than radiation. For $w>1/3$ especially, the ratio of magnetic-to-background energy density actually grows with time, which may possibly enable an initially subdominant PMF to become dynamically relevant unless the duration of this epoch is sufficiently short or the initial magnetic fraction is correspondingly suppressed. Any realistic analysis of PMF-sourced metric perturbations must therefore combine the sourcing calculation with a transparent backreaction criterion.

The majority of existing analyses of PMF-sourced scalar perturbations have focused on the conventional RD limit, where the magnetic anisotropic stress is conveniently normalized by the radiation energy density and therefore becomes effectively time independent prior to neutrino decoupling \cite{ShawLewis2010}. In that limit, one finds the well-known logarithmic growth of the comoving curvature perturbation on superhorizon scales. While invaluable, the RD treatment does not capture the full range of physically plausible post-inflationary histories. In particular, it obscures the way in which a finite reheating era with general constant $w$ changes the temporal behavior of the PMF source, the scalar transfer functions, and the resulting tensor kernel. Likewise, most studies of SIGWs work with scalar sources that are primordial rather than magnetically generated, and therefore do not address the interplay between PMF anisotropic stress, post-inflationary dynamics, and induced tensor production. Establishing this connection is important if one wishes to assess whether a PMF can generate an observable signal in future detectors through the scalar-induced channel, and under what conditions that signal competes with or remains subdominant to the direct first-order magnetic tensor contribution.

The purpose of this work is to develop precisely such a framework, where we focus on scalar metric perturbations sourced by a primordial magnetic field generated at the end of inflation and evolving through a finite post-inflationary epoch characterized by a constant EoS parameter $w$. Working in a gauge-invariant framework, we first derive the sourced evolution of the scalar Bardeen gravitational potentials in a general barotropic background and show explicitly how the standard RD behavior is recovered as the special case $w=1/3$. We then construct the corresponding transfer functions for the PMF-sourced Bardeen potentials and the comoving curvature perturbation, paying particular attention to their dependence on the magnetic switch-on time, the mode wavelength, and the reheating duration. Since the PMF source term is controlled by the anisotropic-stress spectrum, we also derive the associated two-point function for a power-law magnetic field spectrum and identify the parameter space consistent with the absence of significant backreaction on the background expansion. Equipped with these first-order scalar solutions, we next compute the gravitational-wave background induced at second order by the PMF-generated scalar perturbations and compare it with the direct PMF-sourced tensor spectrum for identical choices of underlying magnetic and post-inflationary parameters. Understanding which channel dominates in different parts of parameter space is essential if GW searches are to be used as a diagnostic of primordial magnetogenesis rather than merely as a bound on an effective stochastic background.

Our analysis is thus motivated by three broader considerations. First, the increasing reach of both ground-based and space-based GW detectors spanning decades in frequency domain, \emph{e.g.}, pulsar timing arrays (PTA) \cite{Lentati:2015qwp,Shannon:2015ect,Arzoumanian:2015liz,Qin:2018yhy}, LISA \cite{Audley:2017drz}, Taiji \cite{Guo:2018npi}, Tianqin \cite{Luo:2015ght}, DECIGO \cite{Seto:2001qf,Yagi:2011wg,DECIGO2011}, AION/MAGIS \cite{Badurina:2019hst}, and ET~\cite{ET2010,Maggiore:2019uih}, creates a timely opportunity to connect nonstandard post-inflationary histories to observables on scales many orders of magnitude smaller than those tested by the CMB. Second, PMFs provide a theoretically well-motivated example of a stochastic source that can alter the scalar sector after inflation, thereby challenging the widespread assumption that the relevant scalar perturbations for induced gravitational waves are exhausted by the primordial curvature mode. Third, the analysis serves as a controlled laboratory for studying how anisotropic stress, reheating dynamics, and backreaction constraints combine in a quantitatively predictive setting. In that sense, the results presented here are relevant not only to PMFs themselves but also more broadly to scenarios in which a subdominant source sector with nontrivial stress-energy tensor survives for an extended period between inflation and radiation domination.

\textit{The paper is organized as follows:} In Sec. \ref{sec:genwscal}, we derive the scalar perturbations sourced by the PMF anisotropic stress, beginning with the familiar RD limit and then generalizing to an arbitrary constant-$w$-dominated epoch. In Sec. \ref{sec:gravwaves}, we compute the spectrum of the PMF anisotropic stress, formulate the perturbativity criteria that limit the allowed parameter space and estimate the spectral abundance of second-order tensor perturbations induced by the PMF-sourced scalar modes \emph{vis-\`{a}-vis} that of the direct first-order magnetic tensor contribution. In particular, we find a narrow and technically interesting niche characterized by a kination-like post-inflationary era, a nearly scale-invariant PMF spectrum and a perturbatively consistent initial magnetic energy fraction that may enable the PMF-sourced second-order SIGW background to dominate over the direct PMF-generated signal. We discuss our key results in Sec. \ref{sec:results}, and conclude in Sec. \ref{sec:conclusion} with a summary of the main implications and possible generalizations to more realistic reheating histories and magnetic-field models.

\medskip

\section{Scalar perturbations sourced by a PMF} \label{sec:genwscal}

We begin by setting the stage for the system of equations governing the dynamics of scalar perturbations when sourced by a PMF. According to Ref. \cite{shaw_lewis_2010}, prior to neutrino decoupling ($\tau_\nu$), PMF anisotropic stress can lead to slow amplification of the comoving curvature perturbation ($\zeta$) on superhorizon scales, which otherwise should remain frozen in the minimal picture. In \cite{shaw_lewis_2010}, this has been studied only for the standard RD era, considering instantaneous reheating at the end of inflation. This also reflects the current status of the topic in the literature. Hence, our first target is to generalize this to the context of an arbitrary post-inflationary epoch having a constant value of the EoS parameter ($0<w<1$) and a finite duration expressed in the number of $e$-folds ($N_{\rm reh}$). 
Throughout this section we use the scalar-perturbed line element
\begin{equation}
    ds^2=a^2(\eta)\left[-(1+2\Psi)d\eta^2+(1-2\Phi)\delta_{ij}dx^idx^j\right]\,,
\end{equation}
and a Fourier convention in which convolution factors are kept explicit whenever they enter the magnetic two-point functions below. The velocity variable $V$ denotes the scalar velocity potential in the usual gauge-invariant decomposition of the momentum density, while the scalar anisotropic-stress amplitude $\Pi$ is defined through
\begin{equation}
    \left(\hat{k}_i\hat{k}_j-\frac{1}{3}\delta_{ij}\right)\delta T^i_{\ j}\equiv p\,\Pi=w\rho\,\Pi\,.
\end{equation}
Thus $\Pi$ is dimensionless; although it is convenient to regard it as normalized to the background energy density, the explicit factor of $w=p/\rho$ appears in the traceless Einstein equation. Entropy perturbations are parametrized by $\Gamma$ through the non-adiabatic part of the pressure perturbation, and we specialize to a barotropic fluid with $c_s^2=w$ only after writing the gauge-invariant equations.

The starting point for scalar perturbations is the following system of four coupled linear equations for the scalar perturbation variables:
\beq\label{eq:efe1}
    k^2\Phi=-\dfrac{3}{2}\mathcal{H}^2\left[\Delta+3(1+w)\dfrac{\mathcal{H}}{k}V\right]\:,
\eeq
\beq\label{eq:efe2}
    k\left(\Phi'+\mathcal{H}\Psi\right)=\dfrac{3}{2}\mathcal{H}^2(1+w)V\:,
\eeq
\beq\label{eq:efe3a}
    \Phi''+\mathcal{H}\left(\Psi'+2\Phi'\right)+\left(2\mathcal{H}'+\mathcal{H}^2\right)\Psi+\dfrac{k^2}{3}(\Phi-\Psi)=\dfrac{3w}{2}\mathcal{H}^2(\Delta+\Gamma)\:,
\eeq
\beq\label{eq:efe3b}
    k^2(\Phi-\Psi)=3w\mathcal{H}^2\Pi\:.
\eeq
These four equations follow, respectively, from the $00$, $0i$, $ij$-trace, and $ij$-traceless components of the scalar-projected, first-order perturbed Einstein field equations in Fourier space. Here, $\Phi$ and $\Psi$ are the two gauge-invariant scalar Bardeen potentials, $\Delta$ is the dimensionless density contrast, $V$ is the velocity potential, $\Gamma$ is the entropy perturbation, $\Pi$ is the scalar-projected part of the anisotropic stress normalized with the background energy density, and $\mathcal{H}\equiv a'/a$ is the conformal Hubble parameter. Note that the presence of anisotropic stress results in an offset between the two Bardeen potentials $\Phi$ and $\Psi$, as visible from the $ij$-traceless equation. For clarity, let us spell out the reduction to a single equation for $\Phi$. First, Eqs.~\eqref{eq:efe1} and \eqref{eq:efe2} give
\begin{equation}
    V=\frac{2k}{3\mathcal{H}^2(1+w)}\left(\Phi'+\mathcal{H}\Psi\right),\qquad
    \Delta=-\frac{2k^2}{3\mathcal{H}^2}\Phi-3(1+w)\frac{\mathcal{H}}{k}V\,.
\end{equation}
Second, the traceless equation fixes the gravitational slip as
\begin{equation}
    \Psi=\Phi-3w\left(\frac{\mathcal{H}}{k}\right)^2\Pi\,.
\end{equation}
Substituting these relations into the trace equation and using the background identity $\mathcal{H}'=-(1+3w)\mathcal{H}^2/2$ for constant $w$ gives the sourced Bardeen-potential equation below. The terms proportional to $\mathcal{H}\Pi'$ and $\mathcal{H}'\Pi$ arise from differentiating the gravitational-slip relation, while the $k^2\Gamma/2$ term keeps track of the possible entropy source before it is set to zero in the applications considered later. Combined together, these four equations lead to a single second-order ordinary differential equation (ODE) in $\Phi$ that reads
\beq\label{eq:phieq}
    \Phi''+3\mathcal{H}(1+w)\Phi'+3wk^2\Phi=3w\left(\dfrac{\mathcal{H}}{k}\right)^2\left(\dfrac{k^2}{2}\Gamma-\dfrac{k^2}{3}\Pi+\mathcal{H}\Pi'+2\mathcal{H}'\Pi\right)\:,
\eeq
where we have used $c_s^2=w$ for the speed of sound, as appropriate for a barotropic background with a constant EoS parameter.

\subsection{Revisiting the radiation dominated limit} \label{subsec:rdlim}

The RD limit (\emph{i.e.}, $w=1/3$) of Eq. \eqref{eq:phieq} is well-known in the literature \cite{shaw_lewis_2010}. In this section, we provide an outline of its derivation as a special case of our more general setup, before moving on to the case of arbitrary constant $w$. Firstly, we note that the dimensionless energy and anisotropic stress variables associated with the PMF in \cite{shaw_lewis_2010} are given by
\beq\label{eq:pibRD}
    T^0_0=-\rho_\gamma\Delta_B\:,\:\:T^i_j=p_\gamma\left(\Delta_B\delta^i_j+{\Pi_B}^i_j\right)\:,
\eeq
where $T^\mu_\nu$ is the PMF energy-momentum tensor. As $T^\mu_\nu$ dilutes as $a^{-4}$ with cosmic expansion, one ends up with time-independent $\Delta_B$ and $\Pi_B$ by construction in the RD case, with the entire time dependence being factored out in $\rho_\gamma$ and $p_\gamma$, \emph{i.e.}, in the photon energy density and pressure. Normalizing with only the photon components in Eq. \eqref{eq:pibRD} leads to $\Pi=R_\gamma\Pi_B$, with $R_\gamma=\rho_\gamma/\rho_{\rm tot}$, which is required since the total energy density ($\rho_{\rm tot}$) is the sum of the energy densities carried by both photons and neutrinos. With $\Pi'=0$ and neglecting entropy perturbations, Eq. \eqref{eq:phieq} therefore reduces to
\beq\label{eq:phieqRD}
    3x^2\left(x^2\dfrac{d^2\Phi}{dx^2}+4x\dfrac{d\Phi}{dx}\right)+x^4\Phi=-R_\gamma\Pi_B\left(6+x^2\right)\:,
\eeq
in terms of the dimensionless variable $x= k\eta$, where $\mathcal{H}(\eta)=\eta^{-1}$ has been implemented. While the exact general solution to Eq. \eqref{eq:phieqRD} can be obtained analytically, existing studies have largely focused on its superhorizon behavior, i.e., for $x\ll1$. This can be readily obtained by expanding the general solution around $x=0$ and ignoring $\mathcal{O}(x)$ and higher terms, which yields
\beq\label{eq:phieqRDsuphor}
    \Phi(x\ll1)\approx-\dfrac{c_2}{x^3}+\dfrac{R_\gamma\Pi_B}{x^2}-\dfrac{c_2}{6x}+\dfrac{c_1}{9\sqrt{3}}+\dfrac{1}{27}R_\gamma\Pi_B\left(8-6\gamma_E+3\ln3\right)-\dfrac{2}{9}R_\gamma\Pi_B\ln x\:,
\eeq
where $\gamma_E$ is the Euler-Mascheroni constant. While this expansion apparently contains infrared (IR) divergences, it is important to note that $\Phi$ does not represent a true physical observable. The latter is in fact given by the comoving curvature perturbation ($\zeta$), which, in a general background, is defined in terms of the Bardeen potentials as
\beq\label{eq:zetadef}
    \zeta=\Phi+\dfrac{2}{3(1+w)}\left(\Psi+\dfrac{\Phi'}{\mathcal{H}}\right)\:,
\eeq
for which $\Psi$ can be obtained from $\Phi$ via Eq. \eqref{eq:efe3b}. Constructing $\zeta$ and expanding for small $x$, one automatically gets rid of the $\mathcal{O}(x^{-3})$ and $\mathcal{O}(x^{-2})$ divergences and ends up with
\beq\label{eq:zetaRDsuphor}
    \zeta(x\ll1)\approx-\dfrac{c_2}{6x}+\dfrac{c_1}{6\sqrt{3}}+\dfrac{1}{6}R_\gamma\Pi_B\left(2-2\gamma_E+\ln3\right)-\dfrac{1}{3}R_\gamma\Pi_B\ln x\:.
\eeq
Now, for $\eta<\eta_B$, \emph{i.e.}, before the PMF is generated, $\zeta$ should behave as in the absence of the source term, which implies that at $\eta=\eta_B$, we should be able to match the two different solutions in presence and in absence of $\Pi_B$. Doing this leads to the following condition:
\beq\label{eq:matching}
    -\dfrac{c_2}{6x_B}+\dfrac{c_1}{6\sqrt{3}}+\dfrac{1}{6}R_\gamma\Pi_B\left(2-2\gamma_E+\ln3\right)-\dfrac{1}{3}R_\gamma\Pi_B\ln x_B=\zeta_{\rm prim}\:,
\eeq
where $\zeta_{\rm prim}$ is the (constant) superhorizon curvature perturbation set by inflation. This is an algebraic equation for the two unknown parameters $c_1$ and $c_2$. To close the system and arrive at a solution, we also need to match the Bardeen potential $\Phi(x)$ with its counterpart in absence of PMFs at $x=x_B$.\footnote{Note that the matching condition does not apply to $\zeta'$, which is discontinuous across the $\eta=\eta_B$ slice due to the sudden switching on of the PMF. See also \cite{2010JCAP...05..022B} for a similar matching across a transition involving a suddenly-generated PMF.} For RD, the exact sourceless solution of $\Phi(x)$ is given by \cite{Kohri:2018awv} 
\beq
    \Phi(x)=\dfrac{9}{x^2}\left[\dfrac{\sqrt{3}}{x}\sin\left(\dfrac{x}{\sqrt{3}}\right)-\cos\left(\dfrac{x}{\sqrt{3}}\right)\right]\zeta_{\rm prim}\:.
\eeq
Performing a series expansion of this $\Phi(x)$ for $x\ll1$ (which simply yields $2\zeta_{\rm prim}/3$) and equating it with the series-expanded version of the solution obtained in presence of a PMF at $x=x_B$, we obtain a closed algebraic system of equations for $c_1$ and $c_2$, which may be solved to obtain their expressions in terms of $\zeta_{\rm prim}$, $R_\gamma\Pi_B$, and $x_B$. Further expanding them for $x_B\ll1$ and retaining only the leading order terms, one obtains the familiar expression for the logarithmic growth of $\zeta$ obtained in \cite{shaw_lewis_2010}, which reads
\beq
    \zeta\approx \zeta_{\rm prim}-\dfrac{1}{3}R_\gamma\Pi_B\left[\ln\left(\dfrac{x}{x_B}\right)+\dfrac{1}{2}\left(\dfrac{x_B}{x}-1\right)\right]\:.
\eeq
In the absence of the primordial component $\zeta_{\rm prim}$, the PMF-sourced comoving curvature perturbation in Fourier space is thus proportional to the Fourier amplitude of the magnetic stress-scalar normalized with the background photon energy density, and enhanced logarithmically with time on superhorizon scales. The expression above should be understood as the matched solution for $x\geq x_B$ after the PMF has been switched on. In particular, the familiar logarithmic growth corresponds to the interval $x_B\ll x\ll1$, while the apparently singular term proportional to $x_B/x$ is a transient matching contribution that remains finite at the initial slice $x=x_B$ and becomes subleading once the mode has evolved for a sufficient superhorizon time.

\subsection{General barotropic scenario with constant equation of state} \label{subsec:genw}

Having demonstrated the technique of solving for $\Phi(x)$ and its consistency with the well-known result for the special case $w=1/3$, we now move on to the scenario of our original interest, \emph{i.e.}, an arbitrary constant EoS parameter $w$. In this case, the normalization of the components of the PMF stress-energy has to be done using the background energy density $\rho_w$, which redshifts as $\rho_w\propto a^{-3(1+w)}$. Thus, the dimensionless anisotropic stress $\Pi_B$ is no longer time-independent. It is useful to introduce the shorthand
\begin{equation}
    \beta_w\equiv\frac{2(3w-1)}{1+3w}\,.
\end{equation}
Since $a(\eta)\propto\eta^{2/(1+3w)}$ for a constant-$w$ background, the PMF stress normalized to $\rho_w$ evolves as $a^{3w-1}\propto\eta^{\beta_w}$. Therefore $\beta_w=0$ in RD, $\beta_w>0$ for stiff backgrounds, and $\beta_w<0$ for softer-than-radiation backgrounds.

Following the approach of \cite{shaw_lewis_2010}, we can construct the equation for $\Phi$ in this case as follows:
\begin{align}\label{eq:phieqgenw}
    &\dfrac{(1+3w)^2x^4}{4w}\dfrac{d^2\Phi}{dx^2}+\dfrac{3(1+w)(1+3w)x^3}{2w}\dfrac{d\Phi}{dx}+\left(\dfrac{1+3w}{2}\right)^2x^4\Phi \nonumber \\
    &=\Pi_B^{(0)}\left(\dfrac{x}{x_B}\right)^{\frac{2(3w-1)}{3w+1}}\left[x^2+\dfrac{24}{(3w+1)^2}\right]\:,
\end{align}
where we have suppressed the momentum argument $\boldsymbol{k}$ for $\Phi$ and $\Pi_B^{(0)}$ for notational convenience (these will be restored later when necessary). With the scalar-projection convention adopted in Eq.~\eqref{eq:efe3b}, the RD variable used above is recovered through the identification
\begin{equation}
    \Pi_B^{(0)}\big|_{w=1/3}=-R_\gamma\Pi_B^{\rm (SL)}\,,
\end{equation}
where $\Pi_B^{\rm (SL)}$ denotes the magnetic anisotropic-stress scalar in the notation of Ref.~\cite{shaw_lewis_2010}. With this identification, Eq.~\eqref{eq:phieqgenw} reduces exactly to Eq.~\eqref{eq:phieqRD} in the limit $w\to1/3$. The minus sign is purely conventional and follows from the scalar projection chosen for the gravitational-slip relation; all spectra below depend on the corresponding two-point function of $\Pi_B^{(0)}$ and are therefore insensitive to this sign convention. In deriving Eq. \eqref{eq:phieqgenw}, we have used $\Pi^i_{Bj}=T^i_j/\rho_w\propto a^{3w-1}\propto\eta^\frac{2(3w-1)}{1+3w}$, which implies $\Pi_B(\eta)=\Pi_B^{(0)}\times(\eta/\eta_B)^\frac{2(3w-1)}{1+3w}$. Thus, it is evident that $w=1/3$ acts as a marginal case between a temporally growing source term ($w>1/3$) and decaying source term ($w<1/3$). The general analytic solution to Eq. \eqref{eq:phieqgenw} for arbitrary $w$ is quite complicated and involves integrals over special functions. One may solve Eq. \eqref{eq:phieqgenw} exactly by matching this general solution with the solution for its sourceless counterpart at $x=x_B$, as in the earlier special case with $w=1/3$. The sourceless solution for $x\leq x_B$ is well-known, and given by \cite{Domenech:2019quo} as
\beq
    \Phi(x)=2^\alpha\Gamma(1+\alpha)\dfrac{3(1+w)}{5+3w}\zeta_{\rm prim}\left(\sqrt{w}x\right)^{-\alpha}J_\alpha(\sqrt{w}x)\:,
\eeq
where $\alpha\equiv\frac{5+3w}{2(1+3w)}$ and $\zeta_{\rm prim}$ is the primordial curvature perturbation set by vacuum fluctuations during inflation. In this work, we choose to focus exclusively on the PMF-sourced component, which therefore translates to the initial condition $\Phi(x_B)=\zeta(x_B)=0$. Equipped with these initial conditions, one may then fix the two arbitrary constants in the general solution and uniquely obtain the analytic form of the post-inflationary solution $\Phi(\boldsymbol{k},\eta)$ at $\eta>\eta_B$. Importantly, we note that the overall solutions for the magnetically sourced $\Phi(\boldsymbol{k},x)$, $\Psi(\boldsymbol{k},x)$, and $\zeta(\boldsymbol{k},x)$ are proportional to an overall factor of $\Pi_B^{(0)}(\boldsymbol{k})$, where we have explicitly restored the momentum argument. Thus, one may effectively split the Bardeen potentials as $\Phi(\boldsymbol{k},\eta)=\Pi_B^{(0)}(\boldsymbol{k})\phi_k(\eta)$ and $\Psi(\boldsymbol{k},\eta)=\Pi_B^{(0)}(\boldsymbol{k})\psi_k(\eta)$, and similarly the comoving curvature perturbation as $\zeta(\boldsymbol{k},\eta)=\Pi_B^{(0)}(\boldsymbol{k})\xi_k(\eta)$, where $\Pi_B^{(0)}(\boldsymbol{k})$ fixes the mode amplitude while the time-dependent functions act as transfer functions. 
To avoid confusing these transfer functions with the Bardeen potentials themselves, one may equivalently write
\begin{equation}
    T^{(B)}_{\Phi,k}\equiv\phi_k\,,\qquad
    T^{(B)}_{\Psi,k}\equiv\psi_k\,,\qquad
    T^{(B)}_{\zeta,k}\equiv\xi_k\,.
\end{equation}
In the remainder of the paper we keep the compact notation $\phi_k,\psi_k,\xi_k$, but Table~\ref{tab:notation} records the transfer-function interpretation explicitly. The exact analytic forms of $\phi_k(x)$ and $\psi_k(x)$ are provided in Appendix~\ref{sec:appA}.

\begin{table}[t]
\centering
\begin{tabular}{ll}
\hline
Symbol & Meaning \\
\hline
$\phi_k\equiv T^{(B)}_{\Phi,k}$ & Transfer function for the PMF-sourced Bardeen potential $\Phi$ \\
$\psi_k\equiv T^{(B)}_{\Psi,k}$ & Transfer function for the PMF-sourced Bardeen potential $\Psi$ \\
$\xi_k\equiv T^{(B)}_{\zeta,k}$ & Transfer function for the PMF-sourced comoving curvature perturbation $\zeta$ \\
$x \equiv k\eta$ & Dimensionless time variable \\
$x_B \equiv k\eta_B$ & Value of $x$ at the PMF switch-on time $\eta_B$ \\
$x_{\rm reh} \equiv k\eta_{\rm reh}$ & Value of $x$ at the end of reheating \\
$\Pi_B^{(0)}(\boldsymbol{k})$ & PMF anisotropic-stress amplitude for each Fourier mode at the initial time \\
{} & ($\eta=\eta_B$) normalized with the background energy density \\
\hline
\end{tabular}
\caption{\it Notation used for the transfer functions and dimensionless time variables in the PMF-sourced scalar and tensor sectors.}
\label{tab:notation}
\end{table}

\begin{figure*}[!t]
    \centering   
    {\includegraphics[width=0.49\columnwidth]{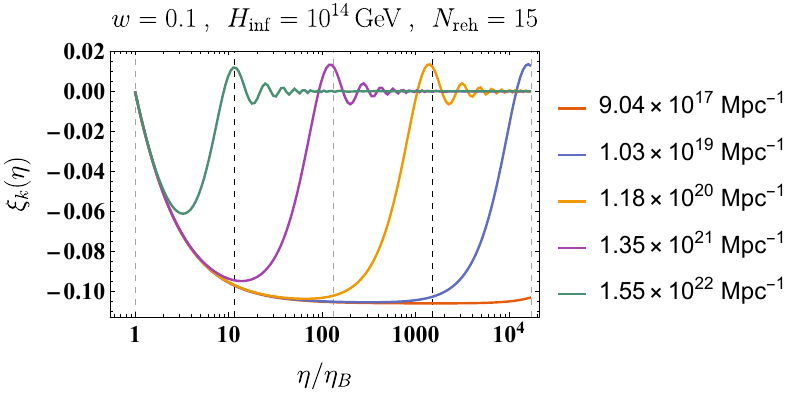}}
    {\includegraphics[width=0.49\columnwidth]{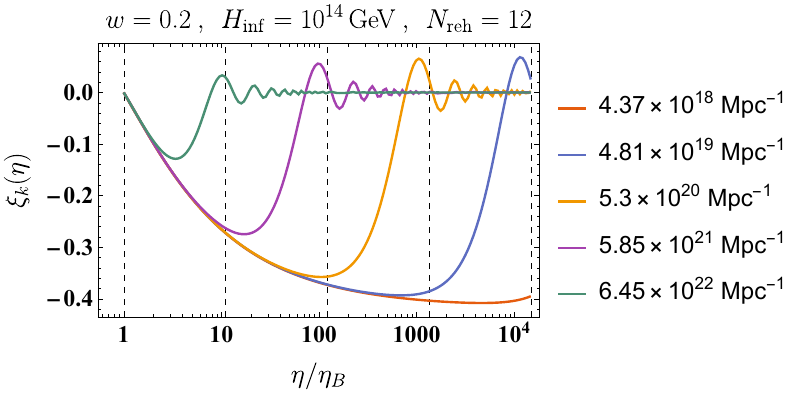}}
    \quad
    {\includegraphics[width=0.49\columnwidth]{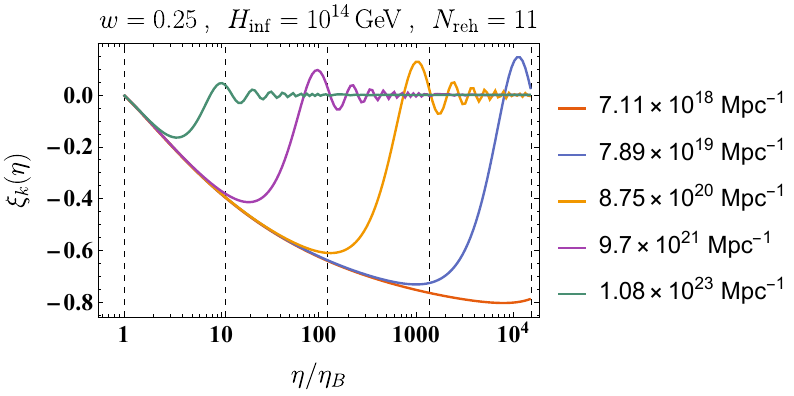}}
    {\includegraphics[width=0.49\columnwidth]{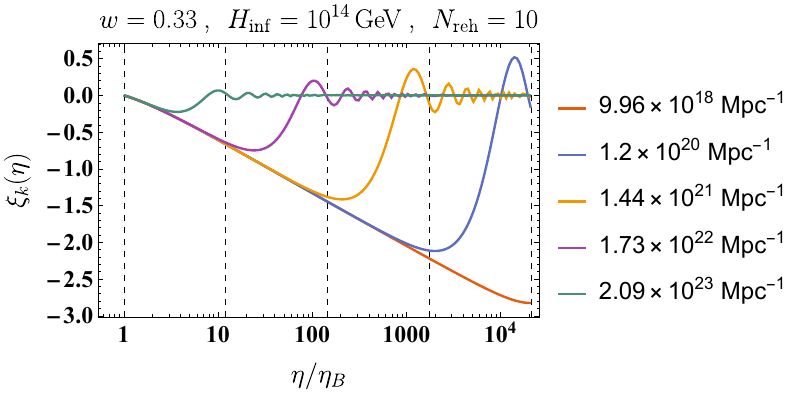}}
    \quad
    {\includegraphics[width=0.49\columnwidth]{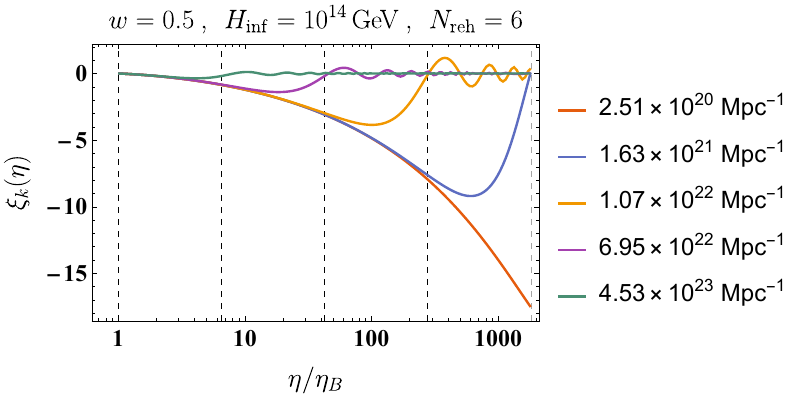}}
    {\includegraphics[width=0.49\columnwidth]{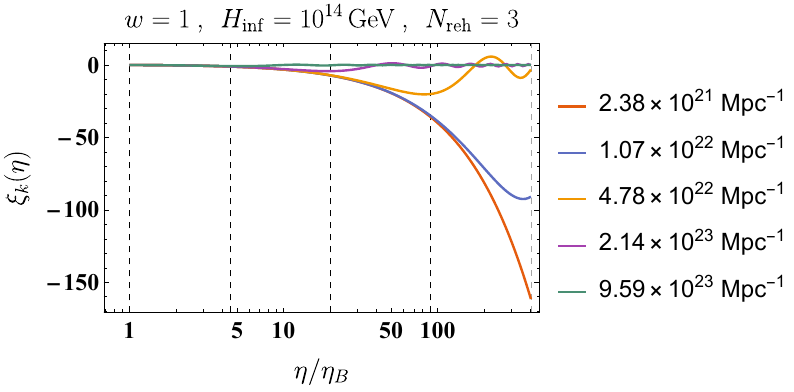}}
    \caption{\it Time evolution of the transfer function $\xi_k(\eta)$ for the PMF-sourced curvature perturbation starting from the initial time $\eta_B=\eta_{\rm inf}$, i.e., the end of inflation, shown in each case for five logarithmically equispaced values of the wavenumber between $k_{\rm inf}$ and $k_{\rm reh}$ (note that the $x$-axis is in log-scale and the $y$-axis is in linear scale). The parameters chosen for each plot are summarized above the corresponding plot. In every scenario, longer wavelength modes stay outside the horizon more time and are comparatively more amplified than shorter modes, which re-enter earlier and subsequently undergo damped oscillations. Furthermore, stiffer backgrounds generically lead to enhanced superhorizon amplification, with the $w=1/3$ scenario (with its logarithmic amplification) acting as a threshold between the concave-up trajectories for $w<1/3$ (slower-than-logarithmic growth) and the concave-down trajectories for $w>1/3$ (faster-than-logarithmic growth).}
    \label{fig:magtransfs_xi}
\end{figure*}

For a qualitative understanding of the characteristic features of the PMF-sourced scalar perturbations, we plot the transfer function $\xi_k(\eta)$ for the comoving curvature perturbation in terms of the reduced variable $\eta/\eta_B$ for different post-inflationary histories in Fig. \ref{fig:magtransfs_xi}. For simplicity, we fix $\eta_B=\eta_{\rm inf}$, \emph{i.e.}, the PMF is assumed to be switched on instantaneously at the end of inflation. While arguably simplistic, this effectively corresponds to a physical scenario in which a classical PMF spectrum originates through the process of inflation on superhorizon scales and assumes a fixed spectral shape as inflation ends, thus leaving its imprints on post-inflationary observables at $\eta>\eta_{\rm inf}$. In each scenario, we show the time evolution of the transfer function corresponding to five logarithmically equispaced values of the wavenumber $k$ lying between $k_{\rm inf}=a_{\rm inf}H_{\rm inf}$ and $k_{\rm reh}=a_{\rm reh}H_{\rm reh}$, \emph{i.e.}, the comoving scales marking the beginning and the end of the post-inflationary era (symbols and subscripts carry their usual meanings). We choose $H_{\rm inf}$ and $N_{\rm reh}$ to be our two fundamental parameters besides $w$, in terms of which, the rest of the quantities introduced above may be written as
\begin{equation} \label{eq:ainf}
    a_{\rm inf}=\left(\dfrac{H_{0}}{H_{\rm eq}}\right)^{2/3}\left(\dfrac{H_{\rm eq}}{H_{\rm inf}}\right)^{1/2}\exp\left[\dfrac{1}{4}(3w-1) N_{\rm reh}\right]\:,
\end{equation}
\begin{equation} \label{eq:areh}
    a_{\rm reh}=a_{\rm inf}\exp(N_{\rm reh})\:,\:\:H_{\rm reh}=H_{\rm inf}\exp\left[-\dfrac{3}{2}(1+w)N_{\rm reh}\right]\:,
\end{equation}
where $H_0\sim10^{-42}$ GeV is the present-day Hubble constant, and $H_{\rm eq}\sim1.5\times10^{-37}$ GeV is its value at matter-radiation equality \cite{Tomberg:2021ajh}. The set of parameter values used for each plot is summarized above the corresponding plot in Fig.~\ref{fig:magtransfs_xi}.

The most prominent feature in Fig. \ref{fig:magtransfs_xi} is clearly the amplification of the PMF-sourced curvature perturbation on superhorizon scales during the post-inflationary epoch. For every post-inflationary scenario under consideration, scalar modes with longer comoving wavelength are observed to spend longer outside the horizon and re-enter later than shorter modes, and are consequently amplified to a larger extent by the PMF source. Furthermore, the transfer function is effectively independent of the wavenumber $k$ deep in the superhorizon regime, as one may notice by the overlap of curves corresponding to different values of $k$ in that regime. This is intuitively expected due to the suppression of spatial gradient terms for $k\ll\mathcal{H}$, which causes each Hubble patch to evolve independently by responding only to its local source irrespective of the perturbation wavelength. After horizon re-entry (marked by a dashed vertical line for each mode), the sourcing mechanism becomes inefficient and the curvature perturbation subsequently starts to decay in an oscillatory fashion. 

Having noted these generic features, it is now instructive to compare among the different possible post-inflationary scenarios demonstrated in Fig. \ref{fig:magtransfs_xi}, which reveals the definitive role played by $w$ in determining the temporal evolution of the PMF-sourced scalar. The RD limit with $w=1/3$, with its associated logarithmic amplification of the superhorizon curvature perturbation, serves as a clear threshold between two qualitatively different branches. For softer values $w<1/3$, the concave-up nature of $\xi_k(\eta)$ signifies a decelerating slower-than-logarithmic growth rate for the curvature perturbation up to horizon re-entry. For stiffer values $w>1/3$, the situation is reversed, with the concave-down shape of $\xi_k(\eta)$ implying an accelerating faster-than-logarithmic growth rate in the superhorizon regime. Thus, the barotropic EoS parameter does not merely control background expansion but also directly governs how efficiently the PMF sector sources scalar perturbations. This is understandable based on the time evolution of the normalized PMF scalar anisotropic stress highlighted earlier, namely $\Pi_B(\eta)\propto \eta^{\frac{2(3w-1)}{1+3w}}$, which leads to a temporally growing (decaying) profile of $\Pi_B(\eta)$ for stiff $w>1/3$ (soft $w<1/3$) values of the EoS parameter. Thus, as time progresses, the PMF sector gradually loses its sourcing capacity on superhorizon scales in a background dominated by a fluid with $w<1/3$, whereas it grows stronger with time for $w>1/3$. This feature becomes consequential when the PMF-sourced scalar modes influence cosmic observables at quadratic or higher order, as we shall revisit throughout later sections.

\section{Second-order gravitational waves induced by PMF-sourced scalars} \label{sec:gravwaves}

Having focused on the dynamics of the PMF-sourced scalar modes in the previous sections, we now move on to study a possible detectable imprint left behind by these perturbations in the form of their induced GW signal. 

\subsection{Spectrum of PMF anisotropic stress}

Since the scalar perturbations are sourced at the linear order by the PMF anisotropic stress, the central quantity for subsequent calculations of both first-order scalar and second-order tensor statistics is the two-point function of the anisotropic stress. The equal-time two-point correlator of a PMF-sourced scalar perturbation yields its corresponding power spectrum, which is thus evidently proportional to the two-point correlator of the initial value of the normalized scalar-projected PMF anisotropic stress $\Pi_B^{(0)}(\boldsymbol{k})$, and hence to the isotropic spectrum of $\Pi_B^{(0)}(\boldsymbol{k})$. We first note that $\Pi_B^{(0)}(\boldsymbol{k})$ is defined as
\beq
    \Pi_B^{(0)}(\boldsymbol{k})\equiv \dfrac{1}{\rho_w(\eta_B)}\left(\hat{k}_i\hat{k}_j-\dfrac{1}{3}\delta_{ij}\right)\int \dfrac{d^3k_1}{(2\pi)^{3/2}}\, B_i(\boldsymbol{k}_1,\eta_B)B_j(\boldsymbol{k}-\boldsymbol{k}_1,\eta_B)\:. 
\eeq
Here and below we use the symmetric Fourier convention $X(\boldsymbol{x})=\int d^3 k\,(2\pi)^{-3/2}X(\boldsymbol{k})e^{i\boldsymbol{k}\cdot\boldsymbol{x}}$. With this convention, quadratic real-space products generate the convolution factor $(2\pi)^{-3/2}$ displayed above. A different Fourier normalization only reshuffles these explicit factors and leaves the dimensionless spectra unchanged once the same convention is used consistently. Defining its two-point correlator as $\langle \Pi_B^{(0)}(\boldsymbol{k})\Pi_B^{(0)}(\boldsymbol{k'})\rangle\equiv \delta^{(3)}(\boldsymbol{k}+\boldsymbol{k'})|\widetilde{\Pi}_B^{(0)}(k)|^2$, one obtains the following expression for the isotropic spectrum:
\beq
    |\widetilde{\Pi}_B^{(0)}(k)|^2=\dfrac{2\pi}{\rho_w(\eta_B)^2}\int_{k_{\rm IR}}^{k_{\rm UV}} dk_1\,k_1^2 P_B(k_1,\eta_B)\int\limits_{-1}^{+1}d\mu\:\mathcal{I}(k,k_1,\mu)P_B\!\left(\sqrt{k^2+k_1^2-2kk_1\mu},\eta_B\right)\:,
\eeq
where $\mathcal{I}(k,k_1,\mu)$ encapsulates the angular contribution and is expressed as\footnote{This can be compared directly with the analogous expression derived in \cite{Saga:2020ics}, where only the RD limit $w=1/3$ has been considered.}
\beq
    \mathcal{I}(k,k_1,\mu)=\dfrac{1}{9}\dfrac{k^2(1+\mu^2)+kk_1(2\mu-6\mu^3)+k_1^2(5-12\mu^2+9\mu^4)}{k^2+k_1^2-2kk_1\mu}\:.
\eeq
Assuming a power law PMF power spectrum of the form $P_B(k,\eta_B)=A_Bk^{n_B}/a(\eta_B)^4$, where $A_B$ is the comoving (present-day) normalization constant in terms of the present-day smoothed PMF strength, we fix the normalization by the Gaussian-smoothed field strength on the scale $\lambda_B$,
\begin{equation}
    B_{\lambda_B}^2\equiv \int \frac{d^3k}{(2\pi)^3}\,P_B(k,\eta_0)\,W^2(k\lambda_B)\,,\qquad W(k\lambda_B)=\exp\!\left(-\frac{k^2\lambda_B^2}{2}\right)\,.
\end{equation}
For a pure power law this relation gives the gamma-function normalization appearing below; changing the smoothing window changes only the numerical prefactor relating $A_B$ and $B_{\lambda_B}$, not the subsequent dependence on the ratios of cutoffs and on $n_B$. One may approximate the power spectrum of the normalized scalar-projected PMF stress at the leading order using the analytic results of \cite{Kosowsky:2001xp} as
\beq \label{eq:anisotropicpowspec}
    |\widetilde{\Pi}_B^{(0)}(k)|^2\approx 2\left[\dfrac{4\pi^2\lambda_B^{n_B+3}}{\Gamma\left(\dfrac{n_B+3}{2}\right)}\right]^2\left[\dfrac{B(\eta_B)^2}{\rho_w(\eta_B)}\right]^2\left[\dfrac{n_B k^{2n_B+3}}{(n_B+3)(2n_B+3)}+\dfrac{k_{\rm UV}^{2n_B+3}}{2n_B+3}-\dfrac{k^{n_B}k_{\rm IR}^{n_B+3}}{n_B+3}\right]\:,
\eeq
where $\lambda_B\sim1$ Mpc is the smoothing scale for estimating present-day PMF strengths, and we use $B(\eta_B)^2=B_\lambda^2/a(\eta_B)^4$ with $B_\lambda$ being the present-day PMF strength smoothed across the typical $\lambda\sim1\:\rm Mpc$ length scale.\footnote{It is important to note that the shorthand $B(\eta)^2=B_0^2/a(\eta)^4$ alone does \emph{not} provide a faithful estimate of the magnetic energy density at any arbitrarily early time, but simply serves to define an effective parameter that we use in our analysis to parametrize the dependence of GW generation on the PMF sector. In contrast, the total PMF energy density is obtained by integrating the PMF power spectrum across a finite range of relevant scales, which leads to backreaction bounds sensitive to the expansion history as well as the magnetic spectral index, as we discuss in Sec. \ref{subsec:perturb}. Hence, we explicitly refrain from interpreting this quantity as a true energy ratio at the end of inflation, but label it as the ``magnetic-to-background'' ratio throughout our analysis to keep this distinction clear.}

Since the benchmark spectra considered below use a nearly scale-invariant magnetic index, it is important to keep the cutoff sensitivity explicit. In the limit $n_B\rightarrow -3^+$ the factors proportional to $(n_B+3)^{-1}$ are regulated only by the finite range of magnetic modes. All numerical estimates in this work should therefore be understood with the physical integration interval $k_{\rm IR}\leq k\leq k_{\rm UV}$ specified by the post-inflationary problem, and in Sec.~\ref{sec:results} we take this interval to be bounded by $k_{\rm reh}$ and $k_{\rm inf}$.

\subsection{Modeling the PMF-sourced SIGW spectrum} \label{subsec:magsigw}

Having computed the power spectrum of the PMF anisotropic stress as a key ingredient, we now proceed to compute the GW spectrum induced at the second order by PMF-sourced scalar modes. Second-order tensor perturbations of the metric may be induced by first-order scalar fluctuations according to the sourcing equation
\begin{equation}
    h_\lambda''(\boldsymbol{k},\eta)+2\mathcal{H}(\eta)h_\lambda'(\boldsymbol{k},\eta)+k^2h_\lambda(\boldsymbol{k},\eta)=S_\lambda(\boldsymbol{k},\eta)\:,
\end{equation}
where $\lambda=\pm1$ is the helicity index labeling the tensor mode function. Assuming sound speed $c_s^2=w$ for the barotropic background fluid, the source term may be generically expressed in the helicity basis in terms of both Bardeen potentials following \cite{Baumann:2007zm} as
The expression below is the scalar-scalar part of the second-order tensor source written without imposing $\Phi=\Psi$. Thus the magnetic anisotropic stress affects the induced channel through the unequal transfer functions $\phi_k$ and $\psi_k$, while the direct first-order tensor sourcing by the PMF stress is treated separately in Sec.~\ref{subsec:maggwdirect}. We neglect entropy sources in the first-order scalar sector when evaluating the kernel, and we do not include first-order vector perturbations in the SIGW source; vector-induced channels constitute a different contribution and are not part of the scalar-induced signal studied here.
\begin{align}
    S_\lambda(\boldsymbol{k},\eta)=\:&4\int\dfrac{d^3q}{(2\pi)^{3/2}}e^{ij}_\lambda(\boldsymbol{k})q_iq_j\Pi_B^{(0)}(\boldsymbol{q})\Pi_B^{(0)}(\boldsymbol{k}-\boldsymbol{q})\left[\dfrac{1-3w}{3(1+w)}\phi_q(\eta)\phi_{|\boldsymbol{k}-\boldsymbol{q}|}(\eta)\right. \nonumber \\
    & \left. +\left(1-\dfrac{2q^2}{3\mathcal{H}^2}\right)\psi_q(\eta)\psi_{|\boldsymbol{k}-\boldsymbol{q}|}(\eta)+2\left(1+\dfrac{q^2}{3\mathcal{H}^2}\right)\phi_q(\eta)\psi_{|\boldsymbol{k}-\boldsymbol{q}|}(\eta) \right. \nonumber \\
    &\left. +\dfrac{2(7+3w)}{3(1+w)\mathcal{H}}\phi_q(\eta)\psi'_{|\boldsymbol{k}-\boldsymbol{q}|}(\eta)-\dfrac{2}{\mathcal{H}}\psi_q(\eta)\psi'_{|\boldsymbol{k}-\boldsymbol{q}|}(\eta)+\dfrac{4}{3(1+w)\mathcal{H}^2}\psi'_q(\eta)\psi'_{|\boldsymbol{k}-\boldsymbol{q}|}(\eta) \right]\:,
\end{align}
where one now has $\phi\neq\psi$ due to the anisotropic stress sourced by the magnetic field. The tensor mode function may then be written as a time integral convolving the source over the Green's function for the tensor equation of motion as
\begin{equation}
    h_\lambda(\boldsymbol{k},\eta)=\int\limits_{\eta_i}^\eta d\tilde{\eta}\:G_k(\eta,\tilde{\eta})S_\lambda(\boldsymbol{k},\tilde{\eta})\:,
\end{equation}
where $\eta_i$ is some fixed initial time, and the Green's function is given by \cite{Domenech:2021ztg}
\begin{equation} \label{eq:greensfunction}
    G_k(\eta,\tilde{\eta})=\dfrac{\pi}{2k}\dfrac{(k\tilde{\eta})^{\sigma+1/2}}{(k\eta)^{\sigma-1/2}}\left[J_{\sigma-1/2}(k\tilde{\eta})Y_{\sigma-1/2}(k\eta)-J_{\sigma-1/2}(k\eta)Y_{\sigma-1/2}(k\tilde{\eta})\right]\:,
\end{equation}
with $\sigma=2/(1+3w)$. The equal-time SIGW two-point function is sourced by the unequal-time two-point correlator of the scalar source function and therefore contains a four-point correlator of the scalar-projected magnetic anisotropic stress $\Pi_B^{(0)}(\boldsymbol{k})$. In the present semi-analytic treatment we retain the disconnected Wick contractions of the effective stress field, i.e. we approximate the relevant stress four-point function by products of the stress two-point function defined above. This should be understood as a Gaussian-stress approximation. It is not equivalent to assuming that the underlying magnetic field is Gaussian, because $\Pi_B^{(0)}$ is quadratic in $B_i$ and therefore possesses connected higher-point functions even for a Gaussian PMF. The connected stress trispectrum would correct the overall normalization and possibly the detailed shape of the induced spectrum; its inclusion requires a dedicated higher-order magnetic correlator calculation and lies outside the scope of the present estimate. Under the disconnected approximation, two non-vanishing Wick contractions give identical contributions to the GW two-point function. Moreover, the contraction between the polarization tensor and the convolution momentum vanishes if the two tensor modes carry different helicity indices, and is identical when both helicity indices are either $+1$ or $-1$. Thus, combining all the contributions from the disconnected contractions and the different helicity modes, the total dimensionless SIGW power spectrum at the end of the post-inflationary epoch is found to be (with $\mathcal{H}_1=\mathcal{H}(\eta_1)$)
\begin{align} \label{eq:magsigwspec}
&\mathcal{P}_{\rm SIGW}^{(B)}(k,\eta_{\rm reh})=\dfrac{32k^3}{\pi^2}\int\dfrac{d^3q}{(2\pi)^3}\lvert e^{ij}_+(\boldsymbol{k})q_iq_j\rvert^2|\widetilde{\Pi}_B^{(0)}(q)|^2|\widetilde{\Pi}_B^{(0)}(|\boldsymbol{k}-\boldsymbol{q}|)|^2 \nonumber \\
& \times \left[ \int\limits_{\eta_B}^{\eta_{\rm reh}}d\eta_1G_k(\eta,\eta_1) \left\{\dfrac{1-3w}{3(1+w)}\phi_q(\eta_1)\phi_{|\boldsymbol{k}-\boldsymbol{q}|}(\eta_1)+\left(1-\dfrac{2q^2}{3\mathcal{H}_1^2}\right)\psi_q(\eta_1)\psi_{|\boldsymbol{k}-\boldsymbol{q}|}(\eta_1)\right.\right. \nonumber \\
& \left.\left. +2\left(1+\dfrac{q^2}{3\mathcal{H}_1^2}\right)\phi_q(\eta_1)\psi_{|\boldsymbol{k}-\boldsymbol{q}|}(\eta_1)+\dfrac{2(7+3w)}{3(1+w)\mathcal{H}_1}\phi_q(\eta_1)\psi'_{|\boldsymbol{k}-\boldsymbol{q}|}(\eta_1) \right.\right. \nonumber \\
&\left.\left. -\dfrac{2}{\mathcal{H}_1}\psi_q(\eta_1)\psi'_{|\boldsymbol{k}-\boldsymbol{q}|}(\eta_1)+\dfrac{4}{3(1+w)\mathcal{H}_1^2}\psi'_q(\eta_1)\psi'_{|\boldsymbol{k}-\boldsymbol{q}|}(\eta_1) \right\} \right]^2\:.
\end{align} 
The structure of Eq.~\eqref{eq:magsigwspec} makes it difficult to extract a closed analytic expression for the full tensor kernel once the exact transfer functions $\phi_k(\eta)$ and $\psi_k(\eta)$ obtained in the general constant-$w$ background are inserted. The relevant time integral contains a nontrivial convolution between the tensor Green's function, the unequal-time scalar response, and derivatives of the PMF-sourced Bardeen potentials. For this reason, the most transparent way to interpret the result is to separate the discussion into two physically distinct layers: first, the time-domain behavior encoded in the transfer functions; second, the momentum-space weighting inherited from the PMF anisotropic stress-spectrum. This separation is not merely technical. It is what makes clear how the post-inflationary history and the PMF microphysics leave complementary imprints on the final SIGW signal. All information from the magnetic sector is channeled into $|\widetilde{\Pi}_B^{(0)}(\boldsymbol{q})|^2|\widetilde{\Pi}_B^{(0)}(|\boldsymbol{k}-\boldsymbol{q}|)|^2$, which overall scales with the fourth power of the energy density ratio between the magnetic sector and the background at $\eta=\eta_B$ (\emph{viz.} Eq. \eqref{eq:anisotropicpowspec}). On the other hand, the magnetic spectral index shapes the momentum dependence of each of these terms, thereby influencing the frequency profile of the SIGW power spectrum. As for the time integral, the background equation of state directly impacts the structure of the Green's function of the tensor mode and the transfer functions of the PMF-sourced Bardeen potentials. To wit, stiffer values of $w$ generally lead to greater superhorizon amplification of the PMF-sourced scalar perturbation, as observed in Fig. \ref{fig:magtransfs_xi}. Additionally, it governs the duration of the post-inflationary epoch in terms of conformal time for a given inflationary energy scale and number of $e$-folds elapsed since the end of inflation, thus determining the comoving horizon size at the onset of RD.

We conclude this section with a few comments on the standard computational recipe that we use to arrive at the final results. To numerically evaluate Eq. \eqref{eq:magsigwspec}, we resort to an explicit coordinate basis where $\boldsymbol{k}$ is aligned along the positive $z$-axis and thus is represented as $\boldsymbol{k}=k(0,0,1)$, and $\boldsymbol{q}=q(\sin\theta\cos\phi,\sin\theta\sin\phi,\cos\theta)$ in terms of the conventional polar angles in this coordinate frame. The polarization tensor then has a matrix representation as the outer product $e_\pm^{ij}(\boldsymbol{k})=e_\mp^i(\boldsymbol{k})e_\mp^j(\boldsymbol{k})/\sqrt{2}$, where the polarization vector is represented as $e_\mp^i(\boldsymbol{k})=(\pm1,-i,0)^{\rm T}/\sqrt{2}$. This may be used to compute the polarization contraction term as
\begin{equation}
    |e_+^{ij}(\boldsymbol{k})q_iq_j|^2=\frac{q^4}{8}\sin^4\theta\,,
\end{equation}
with the polarization normalization stated above. Thereafter, for easier handling of the angular structure, it is useful to transform to the dimensionless momentum variables $v=q/k$ and $u=|\boldsymbol{k}-\boldsymbol{q}|/k$. Similarly, one may also make use of the dimensionless time variable $x_1=k\eta_1$ for the time integral. Imposing finite momentum cutoffs $q_{\rm min}$ and $q_{\rm max}$, the dimensionless variables have the integration domains $u\in\left[\max(v_{\rm min},|1-v|),\min(v_{\rm max},1+v)\right]$ and $v\in\left[v_{\rm min},v_{\rm max}\right]$, where $v_{\rm min,max}=q_{\rm min,max}/k$. The precise values of these integration limits are determined by the physical processes under consideration, and are discussed in detail in Sec. \ref{sec:results}.

\subsection{First-order PMF-sourced tensor spectrum} \label{subsec:maggwdirect}

An important question raised by the formalism developed so far is whether the second-order SIGW background generated by PMF-sourced scalar perturbations can become competitive with, or even dominate over, the direct first-order tensor background generated by the PMF anisotropic stress itself. This comparison is essential for two reasons. First, the two signals originate from the same primordial magnetic sector and may therefore coexist in any realistic stochastic-background prediction. Second, although they share a common microscopic origin, they probe different aspects of the underlying physics. The scalar-induced channel proceeds indirectly through the sourced scalar response of the metric encoded in Eq. \eqref{eq:phieqgenw}, and then through the nonlinear tensor kernel of Eq. \eqref{eq:magsigwspec}. On the other hand, the first-order tensor channel is sourced directly by the magnetic anisotropic stress. The two signals should therefore be viewed as complementary observables that weigh the PMF spectrum and the post-inflationary expansion history in qualitatively different ways.

To make this distinction readily apparent and assess the relative amplitude of both signals in a quantitatively robust manner, it is instructive to explicitly model the direct PMF-sourced tensor spectrum. For a benchmark order-of-magnitude estimate, the dimensionless tensor power spectrum generated directly by the PMF anisotropic stress evaluated at the end of reheating may be expressed as \cite{Bhaumik:2025kuj}
\begin{equation}
    \mathcal{P}_{\rm FO}^{(B)}(k,\eta_{\rm reh})=64G^2k^3\left[\int\limits_{\eta_I}^{\eta_{\rm reh}}d\eta_1\dfrac{G_k(\eta_{\rm reh},\eta_1)}{a(\eta_1)^2}\right]^2\int d^3qP_B(q,\eta_0)P_B(|\boldsymbol{k}-\boldsymbol{q}|,\eta_0)\:,
\end{equation}
where $G$ is the gravitational constant, and the subscript ``FO'' signifies that this spectrum corresponds to tensor fluctuations generated at the first order in the PMF anisotropic stress. In the benchmark calculations below the lower limit of the time integral is identified with the PMF switch-on time, $\eta_I=\eta_B=\eta_{\rm inf}$. We keep the symbol $\eta_I$ in this subsection only to make the formula adaptable to scenarios in which the magnetic source becomes active at a different initial time. As before, the convolution integral over the present-day PMF power spectra may be approximated as \cite{Kosowsky:2001xp}
\begin{equation}
    \int d^3qP_B(q,\eta_0)P_B(|\boldsymbol{k}-\boldsymbol{q}|,\eta_0)\approx 4\pi\left[\dfrac{4\pi^2B_0^2\lambda^{n_B+3}}{\Gamma\left(\frac{n_B+3}{2}\right)}\right]^2\left[\dfrac{n_B k^{2n_B+3}}{(n_B+3)(2n_B+3)}+\dfrac{k_{\rm UV}^{2n_B+3}}{2n_B+3}-\dfrac{k^{n_B}k_{\rm IR}^{n_B+3}}{n_B+3}\right]\:,
\end{equation}
which is a close analogue of Eq. \eqref{eq:anisotropicpowspec}. To rewrite this expression using the magnetic-to-background ratio at the end of inflation as a free parameter, it is useful to eliminate the gravitational constant in terms of the Hubble parameter and the background energy density at the end of inflation using the Friedmann equation, following which the dimensionless tensor power spectrum takes the compact form
\begin{equation} \label{eq:magpowspecfo}
    \mathcal{P}_{\rm FO}^{(B)}(k,\eta_{\rm reh})\approx\:2\pi(H_{\rm inf}a_{\rm inf}^2)^4\lvert\widetilde{\Pi}_B^{(0)}(k)\rvert^2\left[\int\limits_{\eta_I}^{\eta_{\rm reh}}d\eta_1\dfrac{G_k(\eta_{\rm reh},\eta_1)}{a(\eta_1)^2}\right]^2\:,
\end{equation}
where we have used $\eta_B=\eta_{\rm inf}$, and hence, $B(\eta_B)=B_0/a_{\rm inf}^2$, and the power spectrum of the normalized PMF stress-scalar at the end of inflation appears as defined in Eq. \eqref{eq:anisotropicpowspec}.

\subsection{Qualitative comparison between the two channels} \label{subsec:qualcomp}

Before proceeding to numerically compute both the first-order and the second-order spectra, it is interesting to first discuss a few defining characteristics of both spectra that qualitatively distinguish them from each other. At the structural level, the distinction is already evident from the way the magnetic source enters Eqs. \eqref{eq:magsigwspec} and \eqref{eq:magpowspecfo}. In the scalar-induced channel, the PMF anisotropic stress first determines the amplitude of the sourced scalar perturbations through the factorization $\Phi(k,\eta)=\Pi_B^{(0)}(k)\,\phi_k(\eta)$ and $\Psi(k,\eta)=\Pi_B^{(0)}(k)\,\psi_k(\eta)$ discussed in Sec.~2.2. The tensor power spectrum in Eq. \eqref{eq:magsigwspec} then depends on the product $|\widetilde\Pi_B^{(0)}(q)|^2\,|\widetilde\Pi_B^{(0)}(|\mathbf{k}-\mathbf{q}|)|^2$, convolved with a time kernel built from the transfer functions $\phi_k$ and $\psi_k$, their derivatives, and the tensor Green's function. The direct first-order tensor signal, by contrast, is sourced by the PMF anisotropic stress without passing through this intermediate scalar response. In physical terms, the induced channel is sensitive not only to the PMF spectrum itself but also to how efficiently the magnetic source is converted into scalar metric perturbations during the post-inflationary epoch. This additional transfer step is precisely what makes the induced signal especially sensitive to the EoS parameter and the duration of the constant-$w$ era.

A direct consequence of this difference is the distinct parametric dependence of the two tensor channels on the initial magnetic energy fraction. The normalized anisotropic stress-spectrum in Eq. \eqref{eq:anisotropicpowspec} scales as the square of the magnetic-to-background ratio at the switch-on time $\eta_B$. Since the induced tensor spectrum in Eq. \eqref{eq:magsigwspec} contains the product of two such factors, its normalization scales parametrically with the fourth power of the initial magnetic-to-background ratio. The direct first-order tensor background, on the other hand, depends on the PMF anisotropic stress only once at the level of its two-point function. As a result, it is expected to scale more mildly with the magnetic fraction than the scalar-induced signal. This immediately suggests an important phenomenological lesson: for a very weak PMF, the direct tensor channel is generically expected to be easier to excite, whereas the scalar-induced channel becomes relevant only when the scalar transfer functions provide substantial amplification to the SIGW signal with the initial magnetic fraction being large enough to overcome its stronger parametric suppression.

A second distinction concerns the role of the post-inflationary background. The direct first-order magnetic tensor spectrum in Eq. \eqref{eq:magpowspecfo} certainly depends on the expansion history through the tensor Green's function and the $a^{-4}$ evolution of the magnetic source. However, the scalar-induced channel contains an additional layer of sensitivity because the PMF must first generate scalar perturbations that evolve according to Eq. \eqref{eq:phieqgenw}, before these scalar modes subsequently source tensors at second order through Eq. \eqref{eq:magsigwspec}. This means that the post-inflationary EoS parameter $w$ enters the induced channel twice: once through the sourced scalar transfer functions and once through the tensor propagation kernel. By contrast, in the direct first-order channel, the dependence on $w$ is more direct and comparatively less filtered by intermediate scalar dynamics. The curvature transfer function shown in Fig. \ref{fig:magtransfs_xi} already demonstrates that the scalar sector acquires nontrivial scale dependence during the constant-$w$ epoch, with long-wavelength modes receiving enhanced sourcing while short-wavelength modes re-enter early and oscillate away. For the parameter ranges considered below, any tensor background built from these sourced scalars is expected to inherit an enhancement toward modes that remain superhorizon for a longer interval; the precise infrared trend, however, is still shaped by the magnetic spectral index, the convolution kernel, and the imposed cutoffs. In contrast, the first-order PMF tensor signal should not be sensitive to the scalar horizon-crossing history, and this difference in temporal processing is one of the most important conceptual distinctions between the two channels.

The comparison is also shaped by the magnetic spectral index $n_B$ and by the ultraviolet and infrared cutoffs entering Eq. \eqref{eq:anisotropicpowspec}. In the scalar-induced channel, the final tensor signal depends on how the momentum convolution in Eq. \eqref{eq:magsigwspec} samples the normalized PMF stress-spectrum over all pairs of source momenta $(\mathbf{q},\mathbf{k}-\mathbf{q})$. Consequently, the frequency profile of the induced background depends not only on the local behavior of the PMF spectrum near $k$, but also on the global structure of the convolution kernel. This can lead to a frequency dependence that is broader and more strongly shaped by mode coupling than in the direct first-order tensor channel, which depends more directly on the PMF stress-spectrum at the scale of interest.

These structural differences suggest a useful qualitative division of parameter space. In regions where the initial magnetic fraction is small and the sourced scalar transfer functions are only modestly amplified, the first-order PMF tensor signal is expected to dominate simply because it is less parametrically suppressed. This situation is likely to arise for relatively soft reheating histories with limited scalar growth or for PMF amplitudes far below the threshold at which backreaction becomes relevant. By contrast, in regions where the sourced scalar sector experiences significant superhorizon amplification $-$ as suggested by Fig \ref{fig:magtransfs_xi} for modes spending a long interval outside the horizon $-$ the scalar-induced tensor channel can become more competitive. Nevertheless, the magnetic-to-background ratio at the initial time cannot be arbitrarily large as it entails the possibility of perturbative breakdown, which we address in the next section.

\subsection{Perturbativity criterion for sustained GW production} \label{subsec:perturb}

Equipped with the dimensionless tensor power spectrum, one may compute the corresponding GW spectral abundance, $\Omega_{\rm GW}$, as a function of both comoving scale and time as \cite{Maggiore:1999vm}
\begin{equation}
    \Omega_{\rm GW}(k,\eta)=\dfrac{1}{12}\left[\dfrac{k}{\mathcal{H}(\eta)}\right]^2\mathcal{P}_{\rm GW}(k,\eta)\:,
\end{equation}
which is strictly valid under the subhorizon approximation $k\gg \mathcal{H}(\eta)$ which is associated with free wave-like propagation of the generated GWs in an expanding background. In our analysis, this approximation is physically justified for the purpose of calculating the spectral abundance of both the SIGW component and the directly sourced tensor component at $\eta=\eta_{\rm reh}$, as our focus lies on the range of modes that are entirely subhorizon by the instant of reheating.

So far, our analysis rests on the assumption that the PMF energy density redshifts as $a^{-4}$, while the PMF sector continuously sources both first-order scalars and tensors, with the former inducing second order tensor modes on their own. In effect, this assumption is valid as long as the total energy density in the induced tensor channels does not exceed that of the PMF sectors, which would lead to a breakdown of the $a^{-4}$ evolution of the PMF energy density and inhibit further energy transfer from the PMF sector to the tensor sectors. This leads to an additional condition that encapsulates the perturbatively small nature of the total tensor sector compared to the PMF sector, which may be formulated at the end of the reheating as
\begin{equation} \label{eq:perturbcon}
    \Omega_{\rm SIGW}^{(B)}(\eta_{\rm reh})+\Omega_{\rm FO}^{(B)}(\eta_{\rm reh})\ll\Omega_B(\eta_{\rm reh})\:,
\end{equation}
where each term denotes the fraction of total energy carried by each sector, and the net contribution $\Omega_{\rm SIGW}^{(B)}(\eta_{\rm reh})$ may be computed from the spectral abundance as 
\begin{equation}
    \Omega_{\rm SIGW}^{(B)}(\eta_{\rm reh})=\int\limits_{k_{\rm min}}^{k_{\rm max}} \dfrac{dk}{k}\:\Omega_{\rm SIGW}^{(B)}(k,\eta_{\rm reh})\:,
\end{equation}
and likewise for $\Omega_{\rm FO}^{(B)}(\eta_{\rm reh})$. On the right hand side, $\Omega_B(\eta_{\rm reh})$ is given by
\begin{equation}
    \Omega_B(\eta_{\rm reh})=\dfrac{\rho_B(\eta_{\rm reh})}{\rho_w(\eta_{\rm reh})}=\dfrac{\rho_B(\eta_{\rm inf})}{\rho_w(\eta_{\rm inf})}\exp\left[(3w-1)N_{\rm reh}\right]\:,
\end{equation}
where the PMF energy density at the end of inflation may be expressed as
\begin{equation}
    \rho_B(\eta_{\rm inf})=\dfrac{1}{2\pi^2a_{\rm inf}^4}\int\limits_{k_{\rm min}}^{k_{\rm max}}dk\:k^2P_B(k,\eta_0)=\dfrac{2B_\lambda^2\left[(k_{\rm inf}\lambda)^{n_B+3}-(k_{\rm reh}\lambda)^{n_B+3}\right]}{(n_B+3)\Gamma\left(\frac{n_B+3}{2}\right)a_{\rm inf}^4}\:,
\end{equation}
where we set $k_{\rm max}=k_{\rm inf}$ and $k_{\rm min}=k_{\rm reh}$ to capture the full range of comoving scales within the horizon at the end of reheating. Normalizing this with the background energy density allows us to express the magnetic energy density parameter at the end of reheating as
\begin{equation} \label{eq:omegaBreh}
    \Omega_B(\eta_{\rm reh})=\dfrac{B(\eta_{\rm inf})^2}{\rho_w(\eta_{\rm inf})}\times\dfrac{2\left[(k_{\rm inf}\lambda)^{n_B+3}-(k_{\rm reh}\lambda)^{n_B+3}\right]}{(n_B+3)\Gamma\left(\frac{n_B+3}{2}\right)}\exp\left[(3w-1)N_{\rm reh}\right]\:.
\end{equation}

\begin{figure*}[!t]
    \centering   
    {\includegraphics[width=0.45\columnwidth]{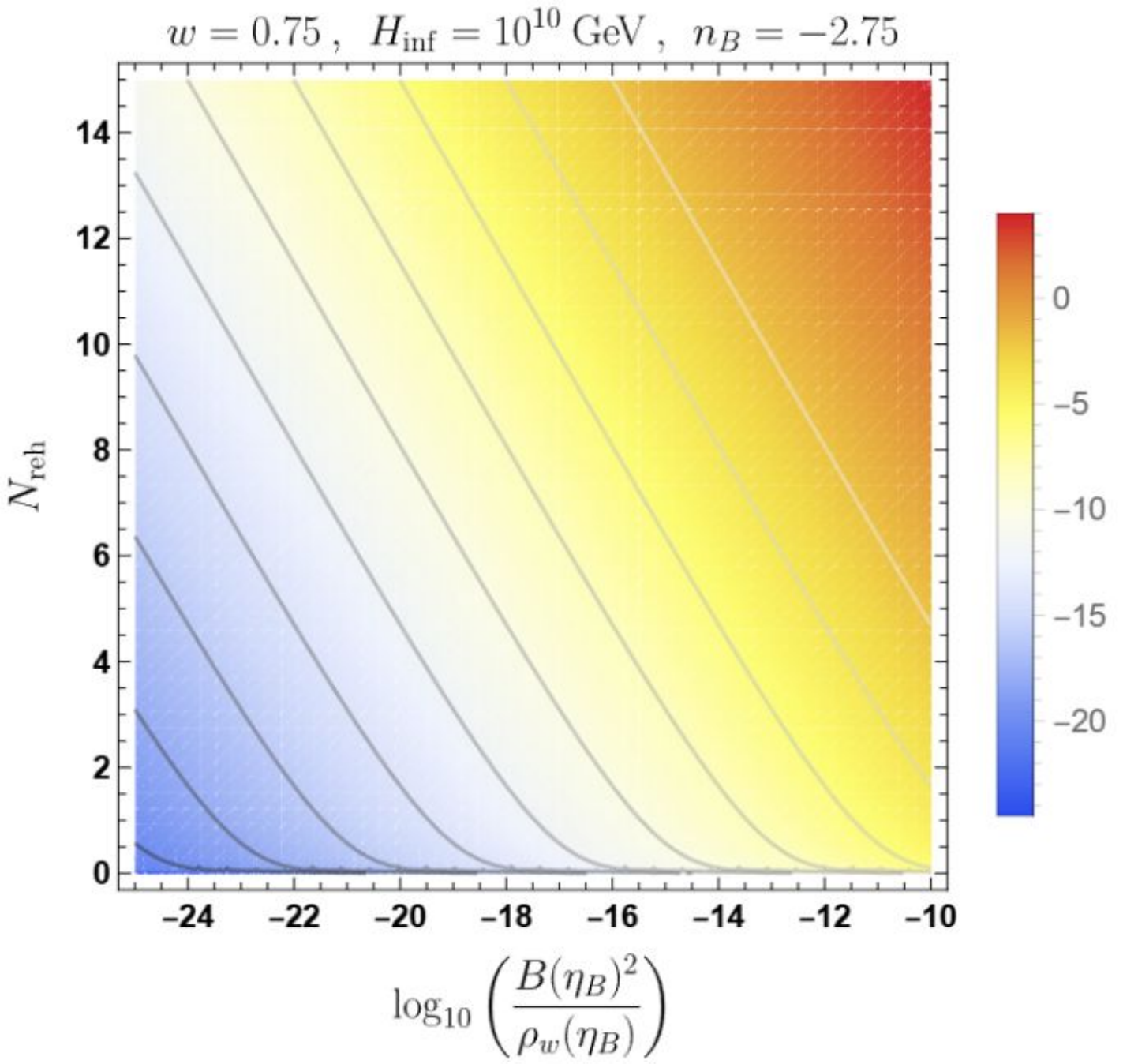}}
    {\includegraphics[width=0.45\columnwidth]{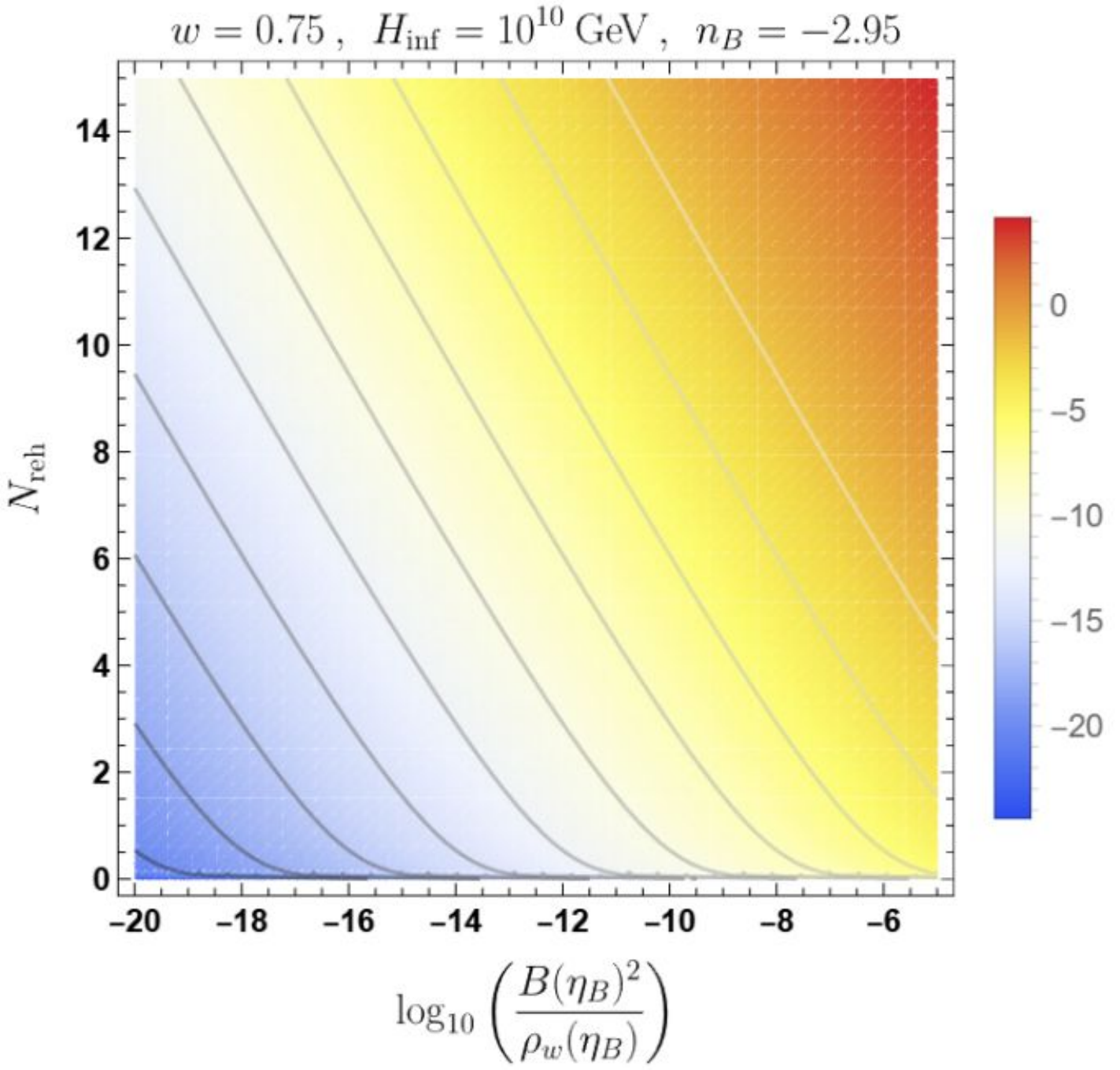}}
    {\includegraphics[width=0.45\columnwidth]{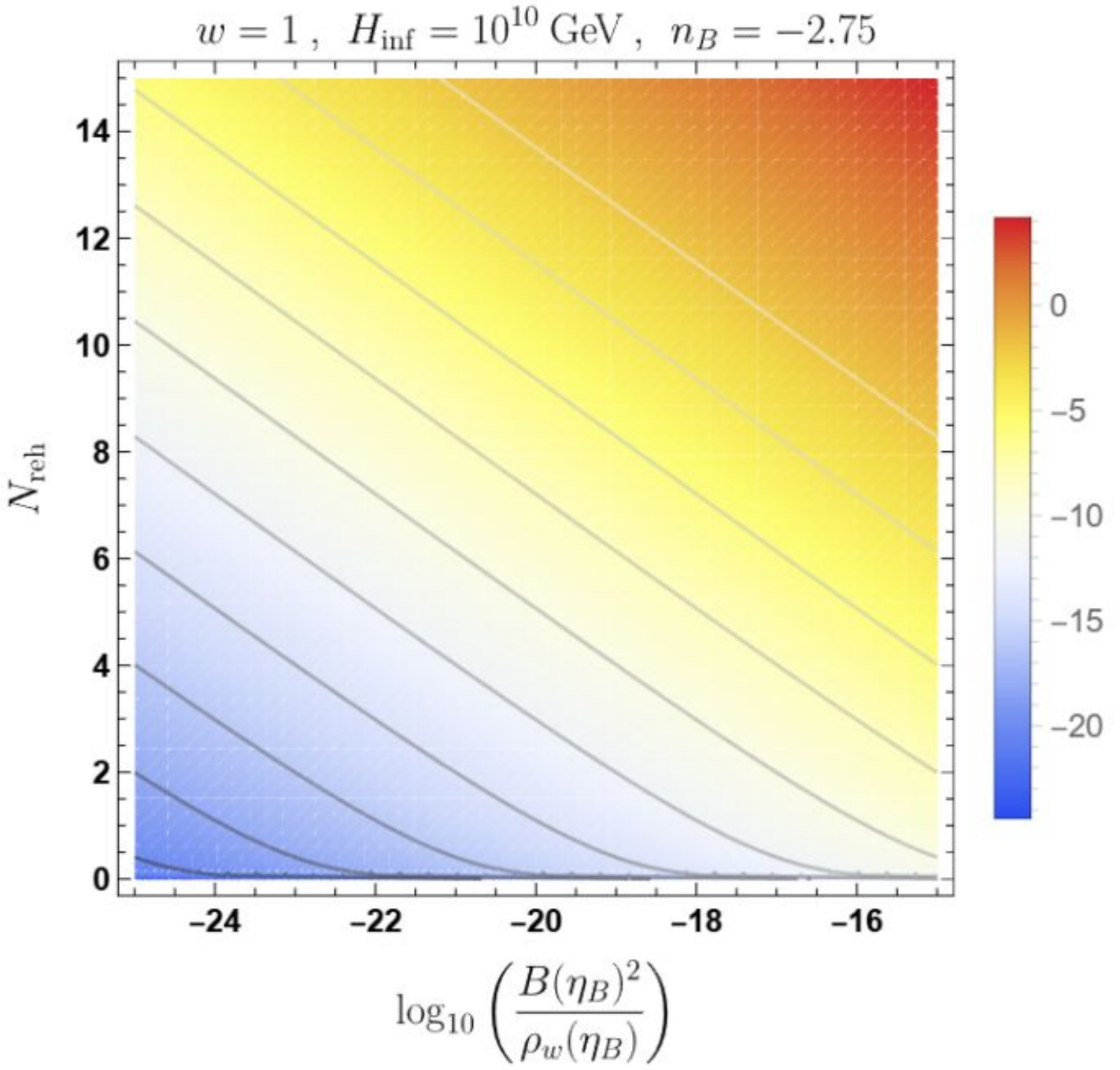}}
    {\includegraphics[width=0.45\columnwidth]{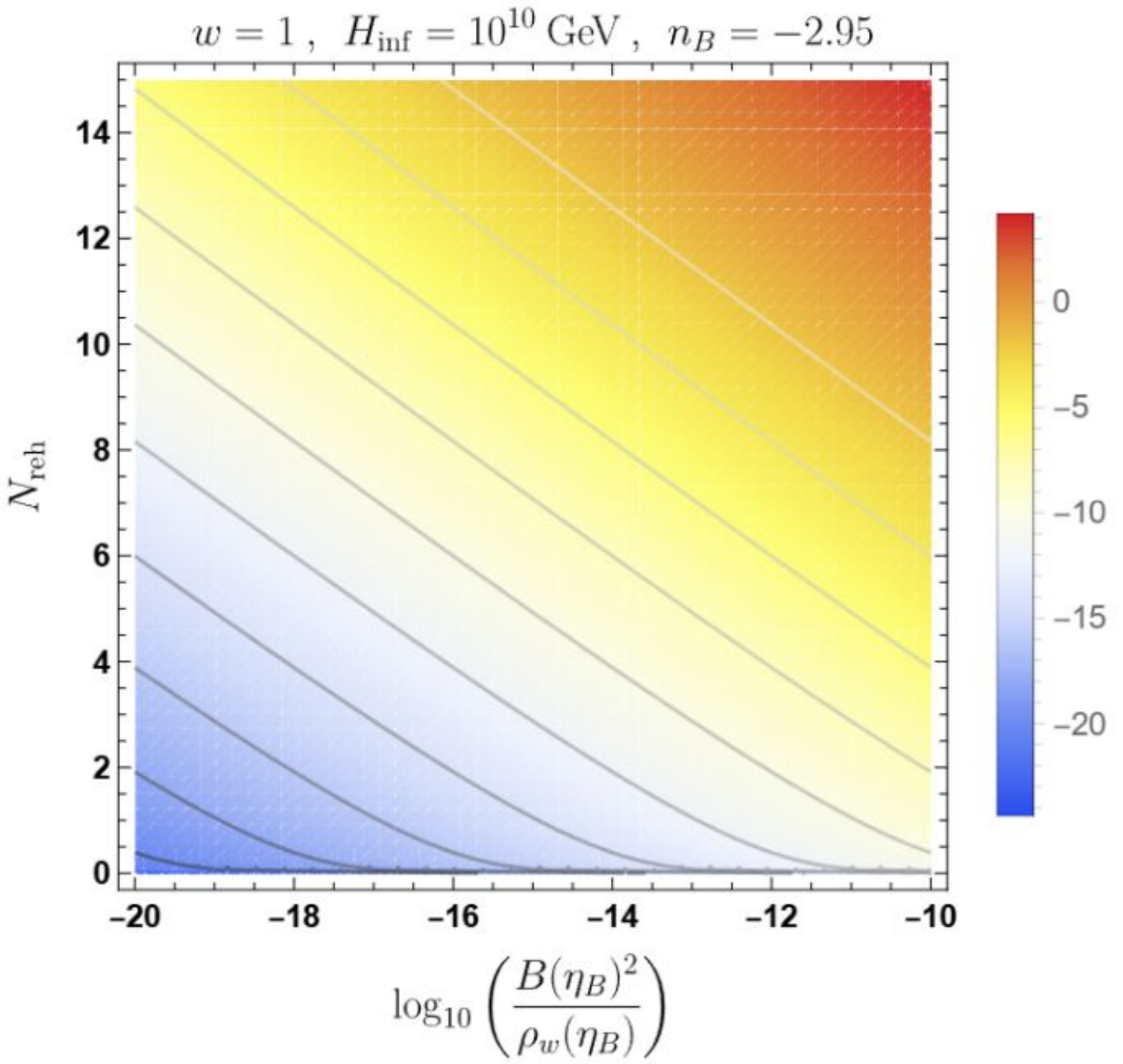}}
    \caption{\it The allowed parameter region satisfying the magnetic backreaction bound $\Omega_B(\eta_{\rm reh})\ll1$ is shown as a function of the initial magnetic-to-background ratio and the duration of the post-inflationary epoch in $e$-folds, for different combinations between the fixed benchmark value of $H_{\rm inf}=10^{10}$ GeV and $n_B=(-2.75,-2.95)$, corresponding to $w=0.75$ (top row) and $w=1$ (bottom row). The magnitude of $\Omega_B(\eta_{\rm reh})$ is colour-coded according to the adjacent panel. The lines correspond to constant $\Omega_B(\eta_{\rm reh})$ contours, with lighter shades corresponding to higher magnitudes.}
    \label{fig:backreactionbound}
\end{figure*}

In addition to the perturbativity condition, it is important that the PMF sector itself remain energetically subdominant compared to the barotropic fluid so that it does not produce significant backreaction and affect background evolution. This essentially requires $\Omega_B(\eta)\ll1$ at $\eta\gtrsim\eta_{\rm inf}$. For a stiff background with $w\geq1/3$, this is ensured automatically if one has $\Omega_B(\eta_{\rm reh})\ll1$, since the energy density fraction in the magnetic sector actually grows with time in such a background. In Fig. \ref{fig:backreactionbound}, we show the allowed parameter region satisfying the magnetic backreaction bound $\Omega_B(\eta_{\rm reh})\ll1$ as a function of the initial magnetic-to-background ratio and the duration of the post-inflationary epoch in $e$-folds, for different combinations between the fixed benchmark value of $H_{\rm inf}=10^{10}$ GeV and $n_B=(-2.75,-2.95)$, corresponding to $w=0.75$ and (top row) and $w=1$ (bottom row). This alone, however, does not guarantee the consistency condition $\Omega_{\rm GW}(\eta_{\rm reh})\ll\Omega_B(\eta_{\rm reh})$. In fact, computing $\Omega_{\rm GW}(\eta_{\rm reh})$ requires evaluating the full GW spectral abundance first, making \emph{a priori} estimates unfeasible. In fact, imposing the full hierarchy of Eq. \eqref{eq:perturbcon} as the baseline criterion severely restricts our available parameter space as far as a dominant PMF-sourced SIGW channel is concerned, as we discuss in the next section. In particular, our analysis suggests that nearly scale-invariant PMF spectra, \emph{i.e.}, with $n_B$ lying close to $-3$, might be particularly capable of simultaneously satisfying the perturbativity condition and leading to a detectably large PMF-sourced stochastic SIGW signal at interferometer scales. 

\section{Results and observational prospects at future detectors} \label{sec:results}

In this section, we provide an explicit comparison between the second-order SIGW and the direct first-order spectra sourced by the stochastic, non-helical PMF under consideration. To highlight the active role of the non-standard post-inflationary background in shaping the GW spectra, we have focused on the range of scales that are fully subhorizon at $\eta=\eta_{\rm reh}$. First, we briefly review a few essential relations that are central to the computation. Once the Universe transits into the standard RD epoch, the $a(\eta)^{-4}$ dilution of the background energy density matches that of the GW energy density, and there is no further temporal evolution of $\Omega_{\rm GW}(k,\eta)$ across the RD epoch. Considering the radiation energy density to scale as $\rho_r = \frac{\pi^2}{30}g_{*\mathrm{\rho}}T_\mathrm{r}^4$, where the temperature of the primordial plasma ($T_\mathrm{r}$) evolves as $T_\mathrm{r}\propto g^{-1/3}_{*\mathrm{S}}a^{-1}$, the GW spectral abundance at the present epoch is given by
\beq\label{Omega_GW_RD_0}
\Omega_\mathrm{GW}(k,\eta_0) = 
\Omega^{(0)}_r\frac{g_{*\mathrm{\rho},\mathrm{*}}}{g_{*\mathrm{\rho},0}}
\left(\frac{g_{*\mathrm{S},\mathrm{0}}}{g_{*\mathrm{S},\mathrm{*}}}\right)^{4/3}
\Omega_{\rm GW}(k,\eta_{\rm reh}),
\eeq
where $g_{*\mathrm{\rho}}$ and $g_{*\mathrm{S}}$ respectively denote the relativistic degrees of freedom of energy and entropy at reheating, and $\Omega^{(0)}_r \sim 10^{-4}$ is the present value of the radiation density parameter. Since reheating has occurred much before BBN, one can show $\frac{g_{*\mathrm{\rho},\mathrm{*}}}{g_{*\mathrm{\rho},0}}\left(\frac{g_{*\mathrm{S},\mathrm{0}}}{g_{*\mathrm{S},\mathrm{*}}}\right)^{4/3}\sim 0.4$. Thus, on sufficiently small scales $k\gtrsim k_{\rm reh}$, the stochastic GW spectral profile at present should essentially trace that at the end of reheating with an overall $\mathcal{O}(10^{-5})$ suppression factor. In the figures and detector comparisons below we plot $h^2\Omega_{\rm GW,0}$ rather than $\Omega_{\rm GW,0}$ itself, where $h$ is the reduced Hubble parameter defined by $H_0=100h\,{\rm km\,s^{-1}\,Mpc^{-1}}$. The conversion is purely notational but is important when comparing with the conventional $\Delta N_{\rm eff}$ bound and with published stochastic-sensitivity curves.

At this point, it is important to note that the onset of RD generates nonlinear magnetohydrodynamic (MHD) turbulence, which becomes important once the characteristic Alfven crossing time becomes comparable to the Hubble time \cite{Banerjee:2004df,Subramanian:2015lua,Sharma:2019jtb}. This is expected to non-trivially affect the subsequent generation of GWs by the PMF sector at both first and second order, since it may simultaneously alter the spectral shape of the effective source term and impact the efficiency of the GW production mechanism itself \cite{Caprini:2006jb,Caprini:2009yp,Brandenburg:2019uzj,Sharma:2022ysf}. An accurate study of this phenomenon requires computationally expensive MHD simulation-based approaches \cite{RoperPol:2019wvy,Brandenburg:2021pdv,Brandenburg:2021bfx,RoperPol:2022iel}, which pushes it outside the analytic scope of the present work. Thus, for the purpose of demonstrating the key results of our current line of inquiry, we restrict ourselves to GW production only during the post-inflationary period, during which the production of charged particles constituting a relativistic plasma is assumed to remain negligible. Neglecting the subsequent MHD era means that the spectra shown here should be interpreted as the contribution generated during the pre-radiation epoch only. Later plasma dynamics, turbulent decorrelation, magnetic damping, neutrino compensation, and changes in the source coherence can modify both the amplitude and the shape of the final spectra, and need not act as a simple overall enhancement. This also compels us to invoke an upper limit of $k_{\rm inf}$ and a lower limit of $k_{\rm reh}$ on the momentum convolution integral, where the former is imposed as a necessary UV-cutoff since the classical treatment of the magnetic power spectrum generated on superhorizon scales during inflation is not generically valid for $k>k_{\rm inf}$, while the latter arises from the fact that modes with $k<k_{\rm reh}$ enter during the RD era when nonlinear MHD effects become important. However, as shown in this section, a large detectable second-order signal may still be obtained for suitable choices of parameters in spite of the aforementioned possibility of underestimation, which may arise from integrating our expressions only across reheating and neglecting PMF-sourced SIGW generation at later times. This is encouraging, but we therefore regard the predicted spectra as controlled pre-radiation contributions rather than as final all-era predictions; a complete treatment of the later plasma era is required before assigning a definitive correction factor to either channel. Hence, instead of the precise quantitative estimates, the broader qualitative takeaway from our current analysis is the finding that, if inflation is succeeded by a brief non-standard post-inflationary epoch driven by a sufficiently stiff background fluid, then it might be possible for the PMF-sourced second-order SIGW to dominate over the first-order PMF-generated GW signal within the detectable thresholds of upcoming GW missions. 

We may now proceed towards an explicit numerical computation of the spectral abundance of both the PMF-sourced SIGW and the direct PMF-sourced tensor modes corresponding to different benchmark values of parameters. This requires consistency with existing bounds based on CMB and BBN observations, which effectively serve to constrain the radiation energy density in the late universe. The latter can be cast in terms of $\Delta N_{\rm eff}=N_{\rm eff}-N_{\rm eff}^{\rm SM}$, which denotes the number of additional degrees of freedom beyond the Standard Model (SM) that are relativistic at the time of recombination, with $N_{\rm eff}^{\rm SM}=3.0440(2)$~\cite{Akita:2020szl,Froustey:2020mcq,Bennett:2020zkv}. Since the effects of adiabatically-generated GWs on the CMB power spectra are similar to those induced by free-streaming dark radiation, it leads to an upper bound on the GW spectral abundance at the present time as follows \cite{Yeh:2022heq}:  
\begin{align}
    h^2\Omega_{\rm GW}(f,\eta_0)\lesssim 5.6\times10^{-6}\;\Delta N_\text{eff} \,. \label{eq:darkrad}
\end{align}
This upper bound is valid for $f\gtrsim2\times10^{-11}$ Hz, \emph{i.e.}, for modes which re-entered the horizon prior to BBN. The \emph{Planck} 2018 + BBN bound of $N_{\rm eff}=2.99\pm 0.17$~\cite{ParticleDataGroup:2022pth} thus implies $\Delta N_{\rm eff}<0.279$ and thus $h^2 \Omega_\text{GW} \lesssim  1.56\times 10^{-6}$ at 95\% C.L., which we denote with the horizontal gray band at the top in the subsequent plots.

\begin{table}[!t]
\centering
\scriptsize
\renewcommand{\arraystretch}{1.35}
\resizebox{\textwidth}{!}{%
\begin{tabular}{c c c c c c c c c}
\hline
Benchmark & $w$ & $H_{\rm inf}$ [GeV] & $N_{\rm reh}$ & $n_B$ & $B(\eta_B)^2/\rho_w(\eta_B)$ & $B_\lambda$ [nG] & $f_{\rm min}$ [Hz] & $f_{\rm max}$ [Hz] \\
\hline
B1 & $1$ & $10^9$ & $8$  & $-2.95$ & $10^{-14.8}$ & $1.1$ & $32.36$ & $1.14\times10^{7}$ \\
B2 & $1$ & $10^5$ & $8$  & $-2.95$ & $10^{-14.8}$ & $1.1$ & $3.24\times10^{-2}$ & $1.14\times10^{5}$ \\
B3 & $1$ & $10^3$ & $8$  & $-2.95$ & $10^{-14.8}$ & $1.1$ & $6.45\times10^{-3}$ & $1.14\times10^{4}$ \\
B4 & $1$ & $10^9$ & $10$ & $-2.95$ & $10^{-18.0}$ & $0.2$ & $1.61$ & $3.08\times10^{7}$ \\
B5 & $1$ & $10^5$ & $10$ & $-2.95$ & $10^{-18.0}$ & $0.2$ & $1.61\times10^{-2}$ & $3.08\times10^{5}$ \\
B6 & $1$ & $10^3$ & $10$ & $-2.95$ & $10^{-18.0}$ & $0.2$ & $1.61\times10^{-3}$ & $3.08\times10^{4}$ \\
\hline
\end{tabular}}
\caption{\it Benchmark parameter choices used in Figs.~\ref{fig:kination_plots_compset1_Nreh8} and \ref{fig:kination_plots_compset2_Nreh10}. The first three rows correspond to the $N_{\rm reh}=8$ spectra in Fig.~\ref{fig:kination_plots_compset1_Nreh8}, while the last three rows correspond to the $N_{\rm reh}=10$ spectra in Fig.~\ref{fig:kination_plots_compset2_Nreh10}. The frequencies are obtained from $f=k/(2\pi)$ with $a_0=1$, using $k_{\rm inf}=a_{\rm inf}H_{\rm inf}$ and $k_{\rm reh}=a_{\rm reh}H_{\rm reh}$ together with Eqs.~\eqref{eq:ainf} and \eqref{eq:areh}.}
\label{tab:benchmarks_fig34}
\end{table}

\begin{figure*}[!t]
    \centering
    {\includegraphics[width=0.85\columnwidth]{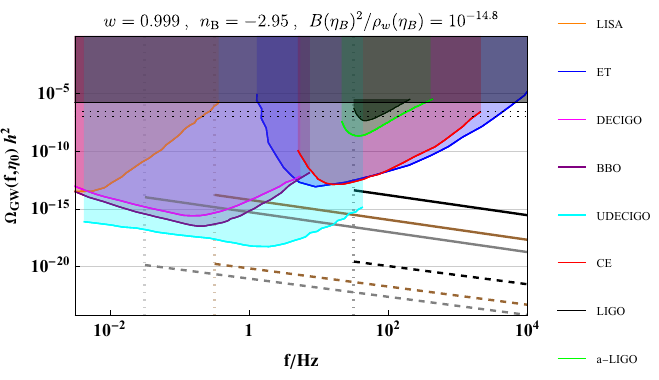}}
    \quad
    \caption{\it Present day spectral abundance envelopes of the PMF-sourced SIGW (solid lines) and the direct PMF-sourced tensor modes (dashed lines) in a kination-like ($w\to1$) post-inflationary background lasting for $N_{\rm reh}=8$ $e$-folds before the onset of standard RD, for three representative values of the Hubble parameter at the end of inflation $H_{\rm inf}=10^{9}$ GeV (black), $10^{5}$ GeV (brown), $10^{3}$ GeV (gray). The plots correspond to a magnetic spectral index $n_B=-2.95$ and a magnetic-to-background ratio at the end of inflation $B(\eta_B)^2/\rho_w(\eta_B)=10^{-14.8}$. The dashed band on top represents the excluded region based on current $\Delta N_{\rm eff}$ constraints from \textit{Planck} 2018 + BBN observations \cite{ParticleDataGroup:2022pth}, while the dotted horizontal lines just below it respectively denote the projected bounds from the upcoming missions CMB-S4 \cite{Abazajian:2019eic} + CMB-Bharat \cite{Adak:2021lbu} and CMB-HD \cite{CMB-HD:2022bsz}. The projected frequency-dependent sensitivity profiles of next-generation interferometric GW detectors are shown alongside to assess detectability. In accordance with our adopted semi-analytic framework, the GW spectra are truncated at $f\sim f_{\rm reh}$ (see text for more detail). This panel corresponds to benchmarks B1-B3 in Table~\ref{tab:benchmarks_fig34}, with the black, brown, and gray curves respectively associated with $H_{\rm inf}=10^9$, $10^5$, and $10^3$ GeV for fixed $N_{\rm reh}=8$, $n_B=-2.95$, and $B(\eta_B)^2/\rho_w(\eta_B)=10^{-14.8}$.}
    \label{fig:kination_plots_compset1_Nreh8}
\end{figure*}

\begin{figure*}[!t]
    \centering
    {\includegraphics[width=0.85\columnwidth]{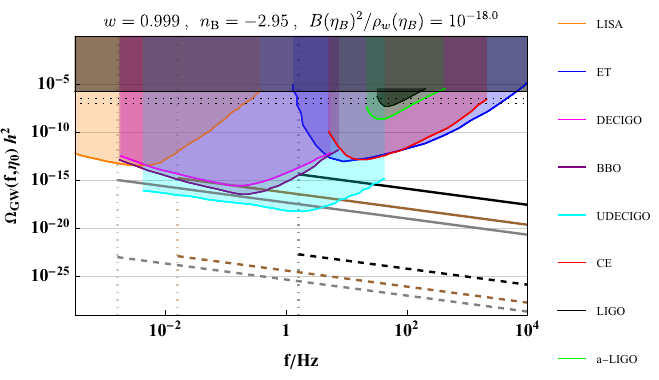}}
    \quad
    \caption{\it Present day spectral abundance envelopes of the PMF-sourced SIGW (solid lines) and the direct PMF-sourced tensor modes (dashed lines) in a kination-like ($w\to1$) post-inflationary background lasting for $N_{\rm reh}=10$ $e$-folds before the onset of standard RD, for three representative values of the Hubble parameter at the end of inflation $H_{\rm inf}=10^{9}$ GeV (black), $10^{5}$ GeV (brown), $10^{3}$ GeV (gray). The plots correspond to a magnetic spectral index $n_B=-2.95$ and a magnetic-to-background ratio at the end of inflation $B(\eta_B)^2/\rho_w(\eta_B)=10^{-18.0}$. Other details are similar to those in Fig. \ref{fig:kination_plots_compset1_Nreh8}. This panel corresponds to benchmarks B4-B6 in Table~\ref{tab:benchmarks_fig34}, with the same color ordering in $H_{\rm inf}$ as in Fig.~\ref{fig:kination_plots_compset1_Nreh8}, but for fixed $N_{\rm reh}=10$, $n_B=-2.95$, and $B(\eta_B)^2/\rho_w(\eta_B)=10^{-18.0}$.}
    \label{fig:kination_plots_compset2_Nreh10}
\end{figure*}

Table~\ref{tab:benchmarks_fig34} collects, in one place, the benchmark inputs used for the two spectral-envelope figures below. The first three benchmarks correspond to the shorter kination-like epoch with $N_{\rm reh}=8$ and a near-boundary magnetic normalization $B(\eta_B)^2/\rho_w(\eta_B)=10^{-14.8}$, while the last three correspond to the longer $N_{\rm reh}=10$ case with the smaller normalization $10^{-18.0}$ required by perturbativity. The table makes clear that lowering $H_{\rm inf}$ shifts both characteristic frequencies downward, moving the same mechanism from ground-based bands toward the deci-Hz and mHz-Hz windows, whereas increasing $N_{\rm reh}$ widens the interval between $f_{\rm reh}$ and $f_{\rm inf}$ and pushes the reheating cutoff to lower frequencies. Thus, Figs.~\ref{fig:kination_plots_compset1_Nreh8} and \ref{fig:kination_plots_compset2_Nreh10} should be read not simply as six independent examples, but as a controlled scan over the inflationary scale and reheating duration within the narrow perturbative region where the PMF-sourced SIGW channel can dominate.

The benchmark table also clarifies how the spectra should scale when moving across the six displayed cases. Since all benchmarks have $w=1$ and $n_B=-2.95$, the differences among the curves are not driven by changes in the stiffness of the background or by changes in the magnetic spectral tilt, but instead by the inflationary scale, the reheating duration, and the magnetic normalization allowed by perturbativity. Increasing $H_{\rm inf}$ at fixed $N_{\rm reh}$ raises both $k_{\rm inf}=a_{\rm inf}H_{\rm inf}$ and $k_{\rm reh}=a_{\rm reh}H_{\rm reh}$ in physical frequency units, and therefore shifts the entire SIGW envelope horizontally toward higher frequencies. This is why the $H_{\rm inf}=10^9\,{\rm GeV}$ benchmarks are most relevant for the high-frequency ground-based range, while the $10^5\,{\rm GeV}$ and $10^3\,{\rm GeV}$ benchmarks move progressively toward the deci-Hz and mHz-Hz bands. By contrast, increasing $N_{\rm reh}$ from 8 to 10 extends the duration over which PMF-sourced scalar modes can grow outside the horizon, but it also forces a smaller initial magnetic-to-background ratio in order to preserve the hierarchy $\Omega_{\rm GW}^{(B)}\ll\Omega_B$. The net result is that the longer reheating benchmarks have a lower reheating cutoff $f_{\rm min}$ and a wider dynamical frequency interval, while their overall normalization is controlled by the smaller value of $B(\eta_B)^2/\rho_w(\eta_B)$. Finally, the nearly scale-invariant choice $n_B=-2.95$ weights the convolution toward a broad range of magnetic modes rather than sharply localizing the signal near the ultraviolet cutoff, so the frequency dependence of the SIGW envelope is governed jointly by the magnetic stress spectrum, the scalar transfer enhancement, and the finite interval $f_{\rm min}<f<f_{\rm max}$ listed in the table.

The detector curves in Figs.~\ref{fig:kination_plots_compset1_Nreh8} and \ref{fig:kination_plots_compset2_Nreh10} are included as reference sensitivity envelopes; the quantitative channel hierarchy should not be inferred only from visual overlap with these curves. For this reason we explicitly quote integrated energy-density ratios in Table~\ref{tab:energycomp}, which provide a cleaner diagnostic of the dominance of the scalar-induced channel over the direct first-order PMF tensor contribution for the displayed benchmarks.

We now proceed to demonstrate our results for a viable region in the underlying parameter space where the PMF-sourced SIGW channel may possibly dominate over the directly sourced GW signal, while satisfying the perturbativity conditions introduced in Sec. \ref{subsec:perturb}. In Fig. \ref{fig:kination_plots_compset1_Nreh8}, we show the best-fit linear envelopes for the present-day PMF-sourced GW spectral abundance induced during a kination-like post-inflationary epoch with $w\to1$ lasting for $N_{\rm reh}=8$ $e$-folds, assuming a magnetic spectral index $n_B=-2.95$ and an initial magnetic-to-background ratio $B(\eta_B)^2/\rho_w(\eta_B)=10^{-14.8}$. These choices satisfy the full consistency hierarchy by obeying both the background backreaction requirement $\Omega_B(\eta_{\rm reh})\ll1$ and the stronger perturbativity condition that the total GW sector remain subdominant to the PMF sector at the end of reheating. The solid curves correspond to the dominant SIGW background, while the dashed curves represent tensor modes sourced directly by the PMF anisotropic stress. The spectra are shown for three representative inflationary Hubble scales, $H_{\rm inf}=10^{9}\,\mathrm{GeV}$, $10^{5}\,\mathrm{GeV}$, and $10^{3}\,\mathrm{GeV}$. As the inflationary scale decreases, the spectra shift systematically towards lower frequencies while largely preserving their overall red-tilted shape, reflecting the corresponding decrease in the characteristic comoving horizon scales at the end of inflation and at the onset of RD. The impact of changing $H_{\rm inf}$ on the peak amplitude is minor for both types of spectra, as it enters only through the dependence of $k_{\rm inf}$ on $H_{\rm inf}$, which regulates the UV cutoff of the momentum convolution in Eq. \eqref{eq:magsigwspec}. Since the short wavelength scalar modes spend less time outside the comoving horizon and are thus amplified to a lesser extent, the contribution of the modes close to the UV end is smaller compared to that of the modes re-entering later. The predicted spectra are compared with the sensitivity curves of current and future GW observatories, illustrating that a significant portion of the parameter space is accessible to next-generation detectors. We show the PMF-sourced SIGW spectrum corresponding to $H_{\rm inf}=10^{9}\,\mathrm{GeV}$ which grazes the projected sensitivities of ET and CE, while those for $H_{\rm inf}=10^{5}\,\mathrm{GeV}$ and $H_{\rm inf}=10^{3}\,\mathrm{GeV}$ fall within the sensitivity bands of DECIGO and BBO. In all cases, the scalar-induced contribution exceeds the directly sourced tensor component by several orders of magnitude throughout the observable frequency range, indicating that the stochastic background is overwhelmingly dominated by the second-order channel. The shaded horizontal band denotes the region excluded by current $\Delta N_{\rm eff}$ constraints, while the dotted horizontal lines indicate the projected sensitivities of future CMB experiments. 

In all the cases, we terminate the spectra at the lower cutoff scale $k\sim k_{\rm reh}$ (indicated by vertical dotted lines in Fig. \ref{fig:kination_plots_compset1_Nreh8}) since our analytic model cannot be reliably extrapolated to larger scales. Since the Green's function of the tensor mode from Eq. \eqref{eq:greensfunction} approaches a $k$-independent limit in the deep superhorizon regime for all values of $w$, dimensional scaling arguments imply that the momentum-dependence of the GW spectra for $k\eta_{\rm reh}\ll1$ should be governed at the leading order by the PMF source terms. Naively, this seems to indicate the generic superhorizon scalings $k^{2n_B+6}$ and $k^{4n_B+12}$ for $\Omega_{\rm FO}^{(B)}(k,\eta_{\rm reh})$ and $\Omega_{\rm SIGW}^{(B)}(k,\eta_{\rm reh})$ respectively. However, as mentioned earlier, an additional layer of complexity arises for these larger scales due to the onset of MHD turbulence in the radiation-dominated epoch, which may alter the dynamics of the effective source on these scales at later times \cite{Caprini:2006jb,Caprini:2009yp,Brandenburg:2019uzj,Sharma:2022ysf}. In turn, this would also modify the evolution of the PMF-sourced scalars both before and after horizon-crossing. Thus, both the first-order and the second-order GW spectra under consideration are expected to get affected at larger scales, as far as their respective present-day spectral abundances are concerned. While causality-based arguments universally appear to predict a blue-tilted GW spectrum in the infrared limit \cite{Cai:2019cdl,Greene:2026one}, the precise spectral shapes of the large-scale GW spectra in our case thus require more careful modeling that accounts for the full range of physics in the RD era. As it falls beyond the scope of this work, we present our results only for $k\gtrsim k_{\rm reh}$, which demonstrates our key qualitative finding that it may be possible for the second-order PMF-sourced GW spectrum to dominate over the first-order PMF-sourced GW spectrum for viable choices of parameters governing the background and the PMF sectors, while retaining observational relevance for small-scale interferometric probes in the coming decades.

In Fig.~\ref{fig:kination_plots_compset2_Nreh10}, the same analysis is repeated for $N_{\rm reh}=10$, corresponding to a slightly longer post-inflationary epoch than in Fig.~\ref{fig:kination_plots_compset1_Nreh8}. Keeping the inflationary scale $H_{\rm inf}$ unchanged shifts $k_{\rm reh}$ towards smaller frequencies as expected, and achieving a sufficiently large SIGW signal amplitude that dominates over the first-order signal requires a smaller initial magnetic-to-background ratio $B(\eta_B)^2/\rho_w(\eta_B)=10^{-18.0}$ which is also consistent with the backreaction bound. Taken together, Figs.~\ref{fig:kination_plots_compset1_Nreh8} and \ref{fig:kination_plots_compset2_Nreh10} show that the relative amplitude of the induced and direct channels is controlled by the interplay between the PMF amplitude and the scalar-transfer enhancement accumulated before reheating. In this sense, the figures do more than merely show benchmark detectability: they expose the mechanism by which a stiff post-inflationary era can reshape both the characteristic frequency and the channel hierarchy. Moreover, for the nearly scale-invariant PMF spectrum under consideration, it is interesting to note that the second-order and the first-order GW spectral envelopes exhibit nearly identical power-law scalings as functions of frequency, with the only difference being their global amplitude. This trend is consistent across all the pairwise cases depicted in Figs. \ref{fig:kination_plots_compset1_Nreh8} and \ref{fig:kination_plots_compset2_Nreh10}. Based on log-log linear fits to the spectral envelopes over the displayed subhorizon interval $f_{\rm reh}<f\lesssim f_{\rm inf}$, we numerically estimate the frequency scalings as
\begin{align}
    &\Omega_{\rm SIGW}^{(B)}(f,\eta_{\rm reh})\propto \left(\dfrac{f}{f_{\rm reh}}\right)^{-0.87\pm0.05}\:, \\ 
    &\Omega_{\rm FO}^{(B)}(f,\eta_{\rm reh})\propto\left(\dfrac{f}{f_{\rm reh}}\right)^{-0.76\pm0.05}\:.
\end{align}
The quoted uncertainty is not a statistical detector error bar; it is a conservative estimate of the spread obtained by varying the fit window within the truncated subhorizon range and should be interpreted as an order-of-magnitude numerical-systematics indicator.

\begin{table}[!t]
\centering
\renewcommand{\arraystretch}{1.6}
\begin{tabular}{|c|c|c|c|}
\hline
 & $H_{\rm inf}=10^9$ GeV & $H_{\rm inf}=10^5$ GeV & $H_{\rm inf}=10^3$ GeV \\
\hline
$N_{\rm reh}=8$ & \begin{tabular}{@{}c@{}}
$\Omega_{\rm SIGW}^{(B)}(\eta_{\rm reh})\sim3.9\times10^{-9}$ \\
$\Omega_B(\eta_{\rm reh})\sim1.0\times10^{-7}$ \\
$\Omega_{\rm SIGW}^{(B)}/\Omega_B\sim3.8\times10^{-2}$
\end{tabular} & \begin{tabular}{@{}c@{}}
$\Omega_{\rm SIGW}^{(B)}(\eta_{\rm reh})\sim1.6\times10^{-9}$ \\
$\Omega_B(\eta_{\rm reh})\sim8.2\times10^{-8}$ \\
$\Omega_{\rm SIGW}^{(B)}/\Omega_B\sim1.9\times10^{-2}$
\end{tabular} & \begin{tabular}{@{}c@{}}
$\Omega_{\rm SIGW}^{(B)}(\eta_{\rm reh})\sim9.8\times10^{-10}$ \\
$\Omega_B(\eta_{\rm reh})\sim7.3\times10^{-8}$ \\
$\Omega_{\rm SIGW}^{(B)}/\Omega_B\sim1.3\times10^{-2}$
\end{tabular} \\
\hline
$N_{\rm reh}=10$ & \begin{tabular}{@{}c@{}}
$\Omega_{\rm SIGW}^{(B)}(\eta_{\rm reh})\sim1.1\times10^{-9}$ \\
$\Omega_B(\eta_{\rm reh})\sim4.2\times10^{-9}$ \\
$\Omega_{\rm SIGW}^{(B)}/\Omega_B\sim2.5\times10^{-1}$
\end{tabular} & \begin{tabular}{@{}c@{}}
$\Omega_{\rm SIGW}^{(B)}(\eta_{\rm reh})\sim4.2\times10^{-10}$ \\
$\Omega_B(\eta_{\rm reh})\sim3.4\times10^{-9}$ \\
$\Omega_{\rm SIGW}^{(B)}/\Omega_B\sim1.2\times10^{-1}$
\end{tabular} & \begin{tabular}{@{}c@{}}
$\Omega_{\rm SIGW}^{(B)}(\eta_{\rm reh})\sim2.6\times10^{-10}$ \\
$\Omega_B(\eta_{\rm reh})\sim3.0\times10^{-9}$ \\
$\Omega_{\rm SIGW}^{(B)}/\Omega_B\sim8.7\times10^{-2}$
\end{tabular} \\
\hline
\end{tabular}
\caption{\it Comparison between the SIGW energy density fraction $\Omega_{\rm SIGW}^{(B)}(\eta_{\rm reh})$ and the magnetic energy density fraction $\Omega_B(\eta_{\rm reh})$ at the end of reheating integrated across the subhorizon range comoving scales, corresponding to the six different scenarios depicted across Figs. \ref{fig:kination_plots_compset1_Nreh8} (upper row in the table with $B(\eta_B)^2/\rho_w(\eta_B)=10^{-14.8}$) and \ref{fig:kination_plots_compset2_Nreh10} (lower row in the table with $B(\eta_B)^2/\rho_w(\eta_B)=10^{-18.0}$) for $n_B=-2.95$. In each case, the induced GW energy density fraction is subdominant to the PMF energy density fraction at the end of reheating and both are significantly smaller than unity, thus satisfying Eq. \eqref{eq:perturbcon} and indicating the validity of our adopted perturbative framework. The third line in each cell gives the ratio $\Omega_{\rm SIGW}^{(B)}/\Omega_B$, making the perturbative hierarchy explicit.}
\label{tab:energycomp}
\end{table}

In addition to the magnetic backreaction bound at the end of reheating, the different scenarios shown in Figs. \ref{fig:kination_plots_compset1_Nreh8} and \ref{fig:kination_plots_compset2_Nreh10} correspond to parts of the parameter space that satisfy the full perturbativity criterion from Eq. \eqref{eq:perturbcon}, which is a requirement for robust GW estimates. In each of the cases shown in Figs. \ref{fig:kination_plots_compset1_Nreh8} and \ref{fig:kination_plots_compset2_Nreh10}, the dominance of the SIGW channel over the first-order one implies that we can approximate the total GW energy density fraction at the end of reheating with $\Omega_{\rm SIGW}^{(B)}(\eta_{\rm reh})$, which may be computed by integrating its spectral abundance over logarithmic $k$-intervals between $k_{\rm reh}$ and $k_{\rm inf}$. On the other hand, the PMF energy density fraction $\Omega_B(\eta_{\rm reh})$ is more straightforward to estimate, based on Eq. \eqref{eq:omegaBreh}. This allows a direct comparison between $\Omega_{\rm SIGW}^{(B)}(\eta_{\rm reh})$ and $\Omega_B(\eta_{\rm reh})$, whose values for the six different scenarios (shown across both plots combined) have been tabulated in Table \ref{tab:energycomp}. For all the cases, $\Omega_{\rm SIGW}^{(B)}(\eta_{\rm reh})/\Omega_B(\eta_{\rm reh})\lesssim10^{-1}$ holds generically, indicating the validity of the perturbative hierarchy that is central to our analysis. It is also important to note that $B(\eta_B)^2/\rho_w(\eta_B)=10^{-14.8}$ in Fig. \ref{fig:kination_plots_compset1_Nreh8} and $B(\eta_B)^2/\rho_w(\eta_B)=10^{-18.0}$ in Fig. \ref{fig:kination_plots_compset2_Nreh10} represent near-extremal upper values of the initial magnetic-to-background ratio for which perturbativity holds, as further increasing the value of this ratio by even half an order-of-magnitude separately in both scenarios quickly results in $\Omega_{\rm SIGW}^{(B)}(\eta_{\rm reh})>\Omega_B(\eta_{\rm reh})$, which seems unphysical as the energy in the perturbatively induced GW sector should not exceed that in the sourcing sector. Thus, further enhancement of the GW spectral amplitude in Figs. \ref{fig:kination_plots_compset1_Nreh8} and \ref{fig:kination_plots_compset2_Nreh10} does not seem possible within our analytic framework by increasing the PMF amplitude. 

That said, we acknowledge this to be an inherently mathematical limitation of our current perturbative framework, since physical intuition does not suggest that any such extremal upper bound $B(\eta_B)^2/\rho_w(\eta_B)\rvert_{\rm max}\ll1$ should in fact exist. In a more realistic scenario, the magnetic-to-background ratio at the initial time needs to satisfy only the PMF backreaction condition $\Omega_B(\eta_{\rm reh})\ll1$ for a $w$-dominated post-inflationary epoch to last for $N_{\rm reh}$ $e$-folds, and GW production from this PMF sector, whether through the first-order or the second-order channel, may proceed by transferring energy from the PMF sector to the GW sector until their energy budgets effectively balance each other, thus attaining an equilibrium between both sectors. Addressing this in a consistent manner requires a framework that would allow us to step beyond the perturbative hierarchy $\Omega_{\rm GW}^{(B)}(\eta)\ll \Omega_B(\eta)$. This requires a self-consistent treatment of the non-adiabatic evolution of the PMF energy density (\emph{i.e.}, modification to the assumed $a^{-4}$ behavior) via a coupled analysis of the perturbed Einstein-Maxwell equations in an expanding background, which lies outside our present scope. Thus, Eq. \eqref{eq:perturbcon} simply highlights the domain of validity of our conclusions in the current work.

Since the magnetic-to-background ratio has been treated as an independent parameter and the background energy density at the end of inflation is given by $\rho_w(\eta_{\rm inf})\sim M_{\rm Pl}^2H_{\rm inf}^2$, it is also important to assess whether the parameter choices in Figs. \ref{fig:kination_plots_compset1_Nreh8} and \ref{fig:kination_plots_compset2_Nreh10} are compatible with the observed $\mathcal{O}(\rm nG)$ upper bound on the present-day PMF strength $B_\lambda$ smoothed on the characteristic $\lambda\sim1\:\rm Mpc$ comoving scale. Using the expression of the scale factor at the end of inflation from Eq. \eqref{eq:ainf}, one may express $B_\lambda$ as
\begin{equation}
    B_\lambda=M_{\rm Pl}H_{\rm eq}\sqrt{\dfrac{B(\eta_B)^2}{\rho_w(\eta_B)}}\left(\dfrac{H_0}{H_{\rm eq}}\right)^{4/3}\exp\left[\dfrac{(3w-1)N_{\rm reh}}{2}\right]\:.
\end{equation}
The parameters used in Fig. \ref{fig:kination_plots_compset1_Nreh8} then correspond to $B_\lambda\sim1.1\:\rm nG$, and those in Fig. \ref{fig:kination_plots_compset2_Nreh10} correspond to $B_\lambda\sim0.2\:\rm nG$. The former should be regarded as a near-boundary benchmark relative to commonly quoted nG-level PMF limits, with the precise level of tension depending on the adopted dataset, smoothing prescription, magnetic spectral index, helicity assumptions, and treatment of post-recombination magnetic effects. The latter is more comfortably within the nG-level range and is therefore less sensitive to the precise observational prior.

On a similar note, what happens for relatively softer EoS parameters $w<1$ is a question worth asking. Our internal computations reveal that reducing the value of the EoS parameter from $w=1$ drastically reduces the size of parameter space for which the following four criteria hold simultaneously: $(i)$ the perturbativity condition is satisfied, $(ii)$ the present-day PMF strength smoothed on $1\:\rm Mpc$ scales satisfies the observed $\mathcal{O}(\rm nG)$ upper bound, $(iii)$ the PMF-sourced SIGW contribution dominates over the first-order contribution, and $(iv)$ the aforesaid SIGW spectrum attains a detectably large amplitude at $f\lesssim10$ kHz which is of interest to future interferometric GW detectors. This requires significant fine-tuning for values of $w$ that are even slightly smaller than $1$, and for even smaller values, there is no parametric region available to simultaneously fulfill all four criteria. Thus, the $w\to1$ post-inflationary epoch complemented with the nearly scale-invariant PMF sector represents a particularly interesting and narrow viable region for which PMF-sourced SIGWs may dominate over direct first-order PMF-sourced tensors within the frequency and amplitude ranges of immediate observable interest.

Before proceeding further, for the benefit of the reader, we once again summarize in this paragraph a few of the important assumptions that underpin our present analysis, and for which the results obtained in this work are valid. We consider a non-helical PMF with a given spectrum to be switched on instantly at the end of inflation $\eta=\eta_{\rm inf}$. Thereafter, we assume inflation to be succeeded by a finite post-inflationary epoch with a constant effective EoS parameter $w$, with the associated sound speed being defined by $c_s^2=w$. The PMF energy density is assumed to redshift universally as $\rho_B\propto a^{-4}$ from the end of inflation onward to the present epoch. In the analysis of the PMF-sourced scalar perturbations in Sec. \ref{sec:genwscal}, entropy perturbations have been neglected. In the subsequent GW analysis which is the focus of Sec. \ref{sec:gravwaves}, only the small-scale GW spectra across the subhorizon range of scales at $\eta=\eta_{\rm reh}$ have been modeled analytically, leaving the detailed MHD simulation-based computation of lower frequency PMF-induced GW spectra to a future analysis.

\subsection*{Projected signal-to-noise ratio (SNR) estimates for GW detectors}

The comparison of the predicted GW spectra against the sensitivity curves of future instruments shown in Figs. \ref{fig:kination_plots_compset1_Nreh8} and \ref{fig:kination_plots_compset2_Nreh10} helps us broadly assess the observability of the signal at the corresponding detectors. For a stronger assessment, we estimate in this section the frequency-integrated signal-to-noise ratio (SNR) of both the first-order and the second-order GW signals at a few of the relevant detectors based on their respective strain noise power spectra $\Omega_{\rm sens}(f)$ \cite{Schmitz:2020syl}.\footnote{The data files for the strain noise power spectra are publicly available at the following repository: \href{https://zenodo.org/records/3689582}{https://zenodo.org/records/3689582}.} Equipped with the theoretically predicted fiducial GW spectrum $\Omega_i^{(B)}(f)$ (where $i$ stands for SIGW/FO) and the noise spectrum $\Omega_{\rm sens}(f)$ for a particular detector, the SNR at the detector may be estimated as
\begin{equation}
    \textrm{SNR}_i=\left[n_{\rm det}T_{\rm obs}\int\limits_{f_{\rm min}}^{f_{\rm max}}df\left(\dfrac{\Omega_i^{(B)}(f)}{\Omega_i^{(B)}(f)+\Omega_{\rm sens}(f)}\right)^2\right]^{1/2}\:,
\end{equation}
where $n_{\rm det}=1$ for an autocorrelation-based detection and $T_{\rm obs}$ is the observation time. In Figs. \ref{fig:kination_plots_compset1_Nreh8} and \ref{fig:kination_plots_compset2_Nreh10}, the fiducial PMF-sourced SIGW spectra shown with solid lines (which dominate over their first-order counterparts depicted with dashed lines) indicate decent observable prospects across the following four detectors: DECIGO, BBO, ET, and CE. In Table \ref{tab:snr}, we tabulate the projected frequency-integrated SNR values at these four instruments (with a total observational time of $T_{\rm obs}\sim4\:\rm years$) for each of the cases shown in the aforementioned figures. As expected, the SNR for the PMF-induced SIGW spectrum ($\rm SNR_{SIGW}$) is much larger than that for the direct PMF-sourced GW spectrum ($\rm SNR_{FO}$). The cases where the spectra do not intersect the plotted integrated stochastic sensitivity curves yield $\rm SNR<10$.

\begin{table}[H]
\centering
\renewcommand{\arraystretch}{1.5}

\begin{tabular}{|c|c|c|c|}
\hline
&
\multicolumn{3}{c|}{SNR estimates for \textbf{Fig.~\ref{fig:kination_plots_compset1_Nreh8}}}
\\
\cline{2-4}
Detector
& $H_{\rm inf}=10^{3}\,\mathrm{GeV}$
& $H_{\rm inf}=10^{5}\,\mathrm{GeV}$
& $H_{\rm inf}=10^{9}\,\mathrm{GeV}$
\\
\hline

DECIGO &
\begin{tabular}{@{}c@{}}
$\textrm{SNR}_{\rm SIGW}\sim50.28$\\
$\textrm{SNR}_{\rm FO}\sim5.25\times10^{-4}$
\end{tabular}
&
\begin{tabular}{@{}c@{}}
$\textrm{SNR}_{\rm SIGW}\sim15.40$\\
$\textrm{SNR}_{\rm FO}\sim3.81\times10^{-4}$
\end{tabular}
&
\begin{tabular}{@{}c@{}}
$\textrm{SNR}_{\rm SIGW}\sim\mathrm{N/A}$\\
$\textrm{SNR}_{\rm FO}\sim\mathrm{N/A}$
\end{tabular}
\\
\hline

BBO &
\begin{tabular}{@{}c@{}}
$\textrm{SNR}_{\rm SIGW}\sim1.54\times10^2$\\
$\textrm{SNR}_{\rm FO}\sim1.28\times10^{-3}$
\end{tabular}
&
\begin{tabular}{@{}c@{}}
$\textrm{SNR}_{\rm SIGW}\sim95.51$\\
$\textrm{SNR}_{\rm FO}\sim2.38\times10^{-3}$
\end{tabular}
&
\begin{tabular}{@{}c@{}}
$\textrm{SNR}_{\rm SIGW}\sim\mathrm{N/A}$\\
$\textrm{SNR}_{\rm FO}\sim\mathrm{N/A}$
\end{tabular}
\\
\hline

ET &
\begin{tabular}{@{}c@{}}
$\textrm{SNR}_{\rm SIGW}\sim3.33\times10^{-4}$\\
$\textrm{SNR}_{\rm FO}\sim2.01\times10^{-9}$
\end{tabular}
&
\begin{tabular}{@{}c@{}}
$\textrm{SNR}_{\rm SIGW}\sim9.61\times10^{-4}$\\
$\textrm{SNR}_{\rm FO}\sim3.74\times10^{-8}$
\end{tabular}
&
\begin{tabular}{@{}c@{}}
$\textrm{SNR}_{\rm SIGW}\sim1.63\times10^{-3}$\\
$\textrm{SNR}_{\rm FO}\sim1.79\times10^{-8}$
\end{tabular}
\\
\hline

CE &
\begin{tabular}{@{}c@{}}
$\textrm{SNR}_{\rm SIGW}\sim7.41\times10^{-4}$\\
$\textrm{SNR}_{\rm FO}\sim8.23\times10^{-9}$
\end{tabular}
&
\begin{tabular}{@{}c@{}}
$\textrm{SNR}_{\rm SIGW}\sim3.92\times10^{-3}$\\
$\textrm{SNR}_{\rm FO}\sim5.51\times10^{-8}$
\end{tabular}
&
\begin{tabular}{@{}c@{}}
$\textrm{SNR}_{\rm SIGW}\sim4.62\times10^{-2}$\\
$\textrm{SNR}_{\rm FO}\sim7.06\times10^{-7}$
\end{tabular}
\\
\hline


&
\multicolumn{3}{c|}{SNR estimates for \textbf{Fig.~\ref{fig:kination_plots_compset2_Nreh10}}}
\\
\cline{2-4}
Detector
& $H_{\rm inf}=10^{3}\,\mathrm{GeV}$
& $H_{\rm inf}=10^{5}\,\mathrm{GeV}$
& $H_{\rm inf}=10^{9}\,\mathrm{GeV}$
\\
\hline

DECIGO &
\begin{tabular}{@{}c@{}}
$\textrm{SNR}_{\rm SIGW}\sim22.04$\\
$\textrm{SNR}_{\rm FO}\sim2.93\times10^{-8}$
\end{tabular}
&
\begin{tabular}{@{}c@{}}
$\textrm{SNR}_{\rm SIGW}\sim1.73\times10^2$\\
$\textrm{SNR}_{\rm FO}\sim1.70\times10^{-7}$
\end{tabular}
&
\begin{tabular}{@{}c@{}}
$\textrm{SNR}_{\rm SIGW}\sim \rm 1.48$\\
$\textrm{SNR}_{\rm FO}\sim \rm 9.48\times10^{-9}$
\end{tabular}
\\
\hline

BBO &
\begin{tabular}{@{}c@{}}
$\textrm{SNR}_{\rm SIGW}\sim47.61$\\
$\textrm{SNR}_{\rm FO}\sim6.50\times10^{-8}$
\end{tabular}
&
\begin{tabular}{@{}c@{}}
$\textrm{SNR}_{\rm SIGW}\sim5.18\times10^2$\\
$\textrm{SNR}_{\rm FO}\sim5.25\times10^{-7}$
\end{tabular}
&
\begin{tabular}{@{}c@{}}
$\textrm{SNR}_{\rm SIGW}\sim \rm 2.14$\\
$\textrm{SNR}_{\rm FO}\sim \rm 1.62\times10^{-8}$
\end{tabular}
\\
\hline

ET &
\begin{tabular}{@{}c@{}}
$\textrm{SNR}_{\rm SIGW}\sim8.28\times10^{-6}$\\
$\textrm{SNR}_{\rm FO}\sim6.77\times10^{-15}$
\end{tabular}
&
\begin{tabular}{@{}c@{}}
$\textrm{SNR}_{\rm SIGW}\sim8.96\times10^{-5}$\\
$\textrm{SNR}_{\rm FO}\sim5.58\times10^{-13}$
\end{tabular}
&
\begin{tabular}{@{}c@{}}
$\textrm{SNR}_{\rm SIGW}\sim6.74\times10^{-2}$\\
$\textrm{SNR}_{\rm FO}\sim1.81\times10^{-11}$
\end{tabular}
\\
\hline

CE &
\begin{tabular}{@{}c@{}}
$\textrm{SNR}_{\rm SIGW}\sim3.78\times10^{-5}$\\
$\textrm{SNR}_{\rm FO}\sim5.94\times10^{-13}$
\end{tabular}
&
\begin{tabular}{@{}c@{}}
$\textrm{SNR}_{\rm SIGW}\sim3.63\times10^{-3}$\\
$\textrm{SNR}_{\rm FO}{\color{revisionorange}\sim}6.79\times10^{-12}$
\end{tabular}
&
\begin{tabular}{@{}c@{}}
$\textrm{SNR}_{\rm SIGW}\sim0.41$\\
$\textrm{SNR}_{\rm FO}\sim1.51\times10^{-10}$
\end{tabular}
\\
\hline

\end{tabular}

\caption{Projected SNR estimates for the dominant PMF-sourced SIGW spectrum ($\rm SNR_{SIGW}$) and the subdominant direct PMF-sourced GW spectrum ($\rm SNR_{FO}$) corresponding to the parameter sets used for Figs. \ref{fig:kination_plots_compset1_Nreh8} and \ref{fig:kination_plots_compset2_Nreh10}, evaluated for observation time $T_{\rm obs}\sim4\:\rm yrs$ with each of the four upcoming interferometric GW detectors DECIGO, BBO, ET, and CE. As usual, we consider SNR $> 10$ to be the standard criterion for robust detectability.}
\label{tab:snr}
\end{table}

\section{Discussion and Conclusion} \label{sec:conclusion}

In this work, we have developed a general framework for studying scalar perturbations sourced by primordial magnetic fields (PMFs) during a finite post-inflationary epoch, and for tracking their imprint on the gravitational wave (GW) background generated at second order. The central result at the level of first-order scalar dynamics is the sourced evolution for the Bardeen potential studied in Sec. \ref{sec:genwscal}, obtained directly from the scalar-projected Einstein equations in a barotropic background with constant equation-of-state (EoS) parameter $w$. The analysis makes transparent that the PMF anisotropic stress enters as a source whose time dependence, normalization, and physical impact are all controlled by the background expansion history. In the radiation-dominated (RD) limit, $w=1/3$, Eq. \eqref{eq:phieqgenw} reduces to its familiar form, and the corresponding superhorizon evolution of the comoving curvature perturbation reproduces the standard logarithmic amplification well-explored in the literature. This recovery of the known RD result provides a first nontrivial consistency check of the formalism and anchors the general constant-$w$ extension to an established limit. The main theoretical advance of the present analysis is that it extends this mechanism beyond radiation domination to a generic reheating-like epoch with constant $w$, where the PMF anisotropic stress can no longer be regarded as effectively time independent when normalized to the background energy density. Physically, this implies that the amplification of PMF-sourced scalar modes is markedly sensitive to the duration and stiffness of the post-inflationary background. This is borne out by the PMF-sourced curvature transfer function shown in Fig. \ref{fig:magtransfs_xi} which exhibits a strong hierarchy between long and short wavelength modes, besides demonstrating that the qualitative character and growth rate of the sourced solution depend appreciably on $w$.

Once the first-order scalar sector is under analytic control, the computation of the PMF-sourced scalar-induced gravitational wave (SIGW) background follows naturally. As shown in Sec. \ref{subsec:magsigw}, the PMF-sourced SIGW signal would therefore be shaped jointly by the post-inflationary history and by the microscopic spectral properties of the magnetic field. This production mechanism is fundamentally different from how the PMF stress-energy directly induces first-order tensor perturbations, as discussed in Sec. \ref{subsec:maggwdirect} and \ref{subsec:qualcomp}. Furthermore, unlike the conventional induced-GW scenario sourced solely by primordial curvature perturbations, here the source sector itself is dynamical and additionally constrained by consistency requirements associated with the magnetic stress-energy tensor. The latter are quantified in terms of the backreaction and perturbativity criteria derived in Sec. \ref{subsec:perturb}, which constitute an essential part of the predictive content of the framework as it determines which regions of the parameter space actually admit a controlled calculation of sourced perturbations.

The current line of inquiry naturally persuades us to investigate whether it might be possible for the second-order SIGW background to become detectably large by dominating over the competing first-order signal, and thus be of interest to next-generation small-scale GW detectors that would collectively span nearly nine decades in frequency space from $10$ $\mu$Hz to $10$ kHz. The findings in Sec. \ref{sec:results} shed light on this prospect by identifying a particular region of parameter space that could make such a scenario feasible $-$ a kination-like post-inflationary epoch ($w\to1$) lasting suitably long ($N_{\rm reh}\sim8-10$) with a nearly scale-invariant ($n_B=-2.95$) non-helical stochastic PMF carrying a sufficiently small fraction of the total energy by the end of reheating ($\Omega_B(\eta_{\rm reh})\sim10^{-8}-10^{-6}$). We emphasize that our analysis does not claim that the dominance of PMF-sourced SIGW over the direct PMF-sourced tensor contribution is generic, but rather that it only arises in a narrow and phenomenologically interesting niche where stiff post-inflationary evolution sufficiently enhances the sourced scalar sector, while remaining compatible with both magnetic backreaction and perturbative GW production bounds as well as the observed present-day $\mathcal{O}(\rm nG)$ upper limit on the smoothed PMF strength. Our results also suggest that this scenario might lead to nearly identical spectral tilts for the SIGW spectrum and the subdominant first-order spectrum, which could therefore serve as a source of degeneracy by obscuring the precise microphysical origin from an observational standpoint. In contrast, post-inflationary scenarios with relatively softer EoS parameters are not typically expected to admit such a parametric region for which the PMF-sourced SIGW channel is the dominant contributor to the total GW energy budget, while being simultaneously consistent with the theoretical perturbativity bounds and remaining observationally interesting.

More broadly, the results of this work reinforce the idea that the post-inflationary expansion history is not a passive background detail but an active part of the GW signal-generation mechanism. In the present case, it governs the time dependence of the magnetic source, the transfer of PMF anisotropic stress into scalar metric perturbations, the horizon-scale timing relevant for both amplification and damping, and the allowed range of magnetically sourced signals consistent with negligible backreaction. From this standpoint, PMFs offer an especially clean example of how post-inflationary physics can leave a nontrivial imprint on observables far removed from the cosmic microwave background (CMB) scales. The formalism presented here is therefore useful not only for PMF phenomenology itself but also as a template for any dynamical source sector that might possibly survive between the end of inflation and the onset of radiation domination.

In conclusion, we have shown that PMF anisotropic stress may source scalar perturbations in a qualitatively richer way than is captured by the standard radiation-dominated treatment, and that a finite post-inflationary constant-EoS epoch may leave significant imprints on both the growth of the sourced scalar sector and the gravitational wave background induced by it at second order. That said, the present work is by no means exhaustive and points in several interesting directions that deserve further investigation. We conclude by highlighting some of them as follows. As already highlighted, a self-consistent calculation of GW-sourcing by a PMF without assuming any \emph{a priori} perturbative hierarchy between the two sectors is expected to be challenging, but merits detailed analysis for more accurate estimates of GW production across wider regions of the parameter space. Furthermore, the study of active generation of GWs during the post-reheating era requires an MHD simulation-based approach, which might provide significant corrections to the analytic estimates presented here. This avenue also includes the detailed compensation of the magnetic anisotropic stress by free-streaming neutrinos after neutrino decoupling, together with a more refined treatment of entropy perturbations and possible departures from the simple barotropic relation $c_s^2=w$. Finally, since the magnetic field sources scalars at the quadratic order, the induced scalar sector is expected to exhibit non-Gaussian features even under the assumption of a Gaussian PMF. The higher-order statistics of the scalar sector could thus provide interesting insights and additional constraints on the details of the PMF sector.\footnote{For instance, this has been studied in an inflationary scenario in \cite{Barnaby:2011vw,Durrer:2024ibi}.} We aim to address some of these questions in future works.

\section*{Acknowledgments} 

Authors thank Debarun Paul, Subhasis Maiti, Deepen Garg, Martin Teuscher, Guillem Dom\`{e}nech, and Misao Sasaki for enlightening discussions. We also acknowledge the use of the APC supercomputing facility. AB thanks CSIR for financial support through Senior Research Fellowship (File no. 09/0093 (13641)/2022-EMR-I). A.G.\ acknowledges the support from the Royal Society, UK, Funding Reference: NIF\ R1\ 253963. TP and AG acknowledge the contribution of the LISA Cosmology Working Group while TP acknowledges financial support from the funding program ``MEDICUS" of the University of Patras.

\appendix

\section{Analytic form of the PMF-sourced Bardeen potential $\Phi(\boldsymbol{k},\eta)$} \label{sec:appA}

It is possible to obtain the general solution to Eq. \eqref{eq:phieqgenw} as
\begin{align} \label{eq:phisolgenapp}
    \Phi(\boldsymbol{k},\eta)=&\left[2(1+3w)x\right]^{-\alpha}\Bigg[c_1(\boldsymbol{k})J_{\alpha}(\sqrt{w}x)+c_2(\boldsymbol{k})Y_{\alpha}(\sqrt{w}x) \nonumber \\
    &\left. +\Pi_B^{(0)}(\boldsymbol{k})\left\{J_\alpha(\sqrt{w}x)\int\limits_1^{x}dy\:\phi_1(y,x_B)+Y_\alpha(\sqrt{w}x)\int\limits_1^{x}dy\:\phi_2(y,x_B)\right\}\right]\:,
\end{align}
where $c_1(\boldsymbol{k})$ and $c_2(\boldsymbol{k})$ are arbitrary constants, and we recall the definitions $\alpha=(5+3w)/[2(1+3w)]$ and $x=k\eta$. The two auxiliary integrands $\phi_{1,2}$ (depending on $x_B=k\eta_B$) are given by
\begin{footnotesize}
    \begin{align}
        \phi_1(y,x_B)=-\dfrac{2^{\alpha+3}\sqrt{w}\left[(1+3w)y\right]^{\alpha-1}\left(\dfrac{y}{x_B}\right)^{3-2\alpha}\left[24+(1+3w)^2y^2\right]Y_\alpha(\sqrt{w}y)}{\left[(1+3w)y\right]^3\left[\left\{J_{\alpha-1}(\sqrt{w}y)-J_{\alpha+1}(\sqrt{w}y)\right\}Y_\alpha(\sqrt{w}y)+\left\{Y_{\alpha+1}(\sqrt{w}y)-Y_{\alpha-1}(\sqrt{w}y)\right\}J_\alpha(\sqrt{w}y)\right]}\:,
    \end{align}
    \begin{align}
        \phi_2(y,x_B)=-\dfrac{2^{\alpha+3}\sqrt{w}\left[(1+3w)y\right]^{\alpha-1}\left(\dfrac{y}{x_B}\right)^{3-2\alpha}\left[24+(1+3w)^2y^2\right]J_\alpha(\sqrt{w}y)}{\left[(1+3w)y\right]^3\left[\left\{J_{\alpha+1}(\sqrt{w}y)-J_{\alpha-1}(\sqrt{w}y)\right\}Y_\alpha(\sqrt{w}y)+\left\{Y_{\alpha-1}(\sqrt{w}y)-Y_{\alpha+1}(\sqrt{w}y)\right\}J_\alpha(\sqrt{w}y)\right]}\:.
    \end{align}
\end{footnotesize}

Based on the expression of $\Phi(\boldsymbol{k},\eta)$ given above, it is straightforward to construct the general solution for $\Psi(\boldsymbol{k},\eta)$, and subsequently, that for $\zeta(\boldsymbol{k},\eta)$, using the traceless $ij$-component of the perturbed Einstein equations. Ignoring any primordial contribution to the scalar sector at the small scales of interest, the initial conditions $\Phi(\boldsymbol{k},\eta_B)=0$ and $\zeta(\boldsymbol{k},\eta_B)=0$ then lead to the explicit forms of $c_{1,2}(\boldsymbol{k})$, which may be plugged back into Eq. \eqref{eq:phisolgenapp} to obtain the exact solution. More explicitly, let $U_1(x)=[2(1+3w)x]^{-\alpha}J_\alpha(\sqrt{w}x)$ and $U_2(x)=[2(1+3w)x]^{-\alpha}Y_\alpha(\sqrt{w}x)$ denote the two homogeneous basis functions, and let $S_\Phi(x,x_B)$ denote the particular solution proportional to $\Pi_B^{(0)}$ in Eq.~\eqref{eq:phisolgenapp}. Defining similarly $U^{\zeta}_{1,2}(x)$ and $S_\zeta(x,x_B)$ by substituting the corresponding $\Phi$ solution into Eq.~\eqref{eq:zetadef}, the matching conditions can be written as the linear system
\begin{equation}
    \begin{pmatrix}
        U_1(x_B) & U_2(x_B) \\
        U^{\zeta}_1(x_B) & U^{\zeta}_2(x_B)
    \end{pmatrix}
    \begin{pmatrix}
        c_1(\boldsymbol{k}) \\
        c_2(\boldsymbol{k})
    \end{pmatrix}
    =-\Pi_B^{(0)}(\boldsymbol{k})
    \begin{pmatrix}
        S_\Phi(x_B,x_B) \\
        S_\zeta(x_B,x_B)
    \end{pmatrix}.
\end{equation}
This form makes the proportionality $c_{1,2}\propto\Pi_B^{(0)}$ manifest and is the most convenient representation for numerical implementation, even when the fully expanded symbolic expressions are lengthy. The latter are too cumbersome to write explicitly, and may be obtained readily via the use of any standard computer algebra software, \emph{e.g.} Mathematica, by following the approach outlined above. Importantly, both $c_1(\boldsymbol{k})$ and $c_2(\boldsymbol{k})$ are overall proportional to $\Pi_B^{(0)}(\boldsymbol{k})$, which thus allows the final decomposition of Eq. \eqref{eq:phisolgenapp} as $\Phi(\boldsymbol{k},\eta)=\Pi_B^{(0)}(\boldsymbol{k})\phi_k(\eta)$, as described in Sec. \ref{subsec:genw}.

\section{Comments on numerical implementation} \label{sec:appB}

For this work, we have followed a hybrid Mathematica + Python-based approach. This allows us to leverage the symbolic capabilities of Mathematica for obtaining analytic mode-matched solutions for the PMF-sourced scalar perturbations in a generic constant-$w$ background, and subsequently use a vectorized numerical pipeline built using Python for swift computation of the second-order PMF-sourced SIGW spectra. The latter step is particularly challenging, as it involves numerically evaluating the derivatives of the complicated scalar transfer functions followed by a full three-dimensional integration over the variables $u=|\boldsymbol{k}-\boldsymbol{q}|/k$, $v=q/k$, and $z_1=k\eta_1$, with the integration domains of the latter two variables typically spanning several decades (\emph{viz.} Eq. \eqref{eq:magsigwspec}). This requires careful implementation of a suitable numerical integration scheme complemented with checks for numerical stability. Here, we briefly discuss a few points in this regard for a clear exposition of our adopted method.

The auxiliary integrands $\phi_{1,2}(y,x_B)$ expressed analytically (\emph{viz.} Sec. \ref{sec:appA}) are used to calculate the integrals $\int\limits_1^{x_B}dy\:\phi_{1,2}(y,x_B)$, which enter the expressions for $c_1(x_B)$ and $c_2(x_B)$ that have been obtained symbolically by mode-matching at $x=x_B$. To compute these integrals, we rely on a log-sampled Simpson's technique and choose $\sim 4\times10^3$ logarithmically equispaced sampling points spanning the decades-long integration domain $y\in[1,x_B]$. Increasing the number of sampling points by roughly one order of magnitude generally results in $\lesssim10\%$ difference in the values of $c_1(x_B)$ and $c_2(x_B)$ which demonstrates convergence for the parameter space considered in this work. For the moderate values of $N_{\rm reh}$ chosen in Sec. \ref{sec:results}, catastrophic cancellation between large-magnitude terms does not play a significant role in the estimates of the final spectra, up to the default 16-digit precision level of the \texttt{numpy} library. Increasing $N_{\rm reh}$ beyond this level typically leads to large and erroneous initial numerical values of $\phi_k(\eta_B)\neq0$ for longer wavelength modes due to cancellation errors, which physically mimics non-zero initial magnitudes of the scalar mode that exist from before the switching-on the PMF. For longer reheating durations, the scalar transfer functions should therefore be evaluated either with arbitrary precision or with analytically rearranged expressions that cancel the large homogeneous and particular pieces before numerical evaluation. We leave such high-precision extensions for future work.

For computing the $z_1$-derivatives of the scalar transfer functions, we have used a forward-difference method with an absolute step size of $\Delta z_1\sim10^{-5}z_1$, which appears sufficient to attain numerical convergence of the final spectra for moderate values of $N_{\rm reh}$ (for $N_{\rm reh}=5$, $\Delta z_1\sim10^{-2}z_1$ leads to the same final order of magnitude as for $\Delta z_1\sim10^{-5}z_1$). Since the limits of the convolution integrals span several decades, the nested 3D numerical integration is subsequently performed using Simpson's method after a suitable exponential transformation of variables, with logarithmically equispaced sampling points across a 3D grid. All multidimensional integrations are carried out in dimensionless variables. When an integration range spans decades, we use logarithmic coordinates and include the corresponding Jacobian factors explicitly. Thus, the Simpson rule is applied on uniform grids in the logarithmic variables rather than on uniform grids in the original variables. This avoids oversampling the ultraviolet end of the integration domain and gives stable weights to both early-time and late-time contributions. The computational cost and runtime thus scale as $\mathcal{O}(N^3)$ for an identical number of sampling points $N$ along each grid axis. For the parameters under consideration, decent convergence is achieved for $N\sim50$ for a reliable order-of-magnitude estimate at each wavenumber. Endpoint and removable-singularity issues are handled explicitly for the boundaries $u=|1-v|$ and $u=1+v$, where the Jacobian-transformed integrand can otherwise be numerically stiff. These checks support the order-of-magnitude spectra reported in Sec.~\ref{sec:results}, but a precision-level forecast would require a dedicated convergence table for every benchmark and detector band. Since the present work aims at controlled order-of-magnitude forecasts rather than percent-level numerical predictions, we regard stability at the level of the quoted spectral envelope and perturbativity hierarchy as the relevant convergence criterion.

\bibliographystyle{JHEP}
\bibliography{main.bib}
\end{document}